\documentclass{article}
\usepackage[utf8]{inputenc}
\usepackage{a4wide}
\usepackage{xcolor}
\usepackage{amsmath}
\usepackage{lscape}
\usepackage[font=small,labelfont=bf]{caption}
\usepackage{graphicx}
\usepackage{subcaption}
\usepackage{comment}
\usepackage{tabularx}
\usepackage{multirow}
\usepackage{algorithmic}
\usepackage[ruled,vlined]{algorithm2e}
\usepackage{amsmath,amssymb,amsfonts,amsthm} 
\usepackage{geometry}
\usepackage{hyperref}
\usepackage{authblk}
\usepackage{ulem}
\usepackage{moreverb}
\usepackage{bm}
\usepackage{siunitx}

{
\newtheorem{theorem}{Theorem}

\newtheorem{assumption}{Assumption}
\theoremstyle{remark}
\newtheorem{remark}{Remark}

}

\renewenvironment{abstract}
  {\small\noindent\textbf{Abstract.}\normalsize\ignorespaces}
  {\par\noindent\ignorespacesafterend}

\makeatletter
\renewcommand{\maketitle}{%
    \noindent\fcolorbox{red}{white}{% Red border, white background
        \parbox{\textwidth}{%
            \color{red}\copyright2025 Elsevier. Accepted for \textit{Journal of Computational Science}. Personal use of this material is permitted. Permission from Elsevier must be obtained for all other uses, in any current or future media, including reprinting/republishing this material for advertising or promotional purposes, creating new collective works, for resale or redistribution to servers or lists, or reuse of any copyrighted component of this work in other works.%
        }%
    }
    \vspace{1em} % Add some vertical space between the box and the title
    \begin{center}%
        {\LARGE \@title \par}% Title
        \vspace{0.5em}% Add some vertical space
        {\large \@author \par}% Author
        \vspace{0.5em}% Add some vertical space
        {\large \@date}% Date
    \end{center}%
    \vspace{1em} % Add some vertical space after the title
}
\makeatother

\begin{document}

\title{Bearing-Distance Flocking with Zone-Based Interactions in Constrained Dynamic Environments}

\author{Hossein B. Jond}
\affil{Department of Cybernetics, Czech Technical University in Prague, Prague, Czechia}

\date{}

\maketitle

\begin{abstract}
This paper presents a novel zone-based flocking control approach suitable for dynamic multi-agent systems (MAS). Inspired by Reynolds behavioral rules for \textit{boids}, flocking behavioral rules with the zones of repulsion, conflict, attraction, and surveillance are introduced. For each agent, using only bearing and distance measurements, behavioral contribution vectors quantify the local separation, local and global flock velocity alignment, local cohesion, obstacle avoidance and boundary conditions, and strategic separation for avoiding alien agents. The control strategy uses the local perception-based behavioral contribution vectors to guide each agent's motion. Additionally, the control strategy incorporates a directionally aware obstacle avoidance mechanism that prioritizes obstacles in the agent's forward path. Simulation results validate the effectiveness of the model in creating flexible, adaptable, and scalable flocking behavior. Asymptotic stability and convergence to a stable flocking configuration for any initial conditions provided the interaction graph is a spanning tree are demonstrated. The flocking model's reliance on locally sensed bearing and distance measurements ensures scalability and robustness, particularly in scenarios where communication is unreliable or resource-intensive. This makes it well-suited for real-world applications demanding seamless operation in highly dynamic and distributed environments.

\end{abstract}

\section{Introduction}
The concept of flocking draws inspiration from the synchronized collective motion observed in birds, fish, and other social animals. In 1987, Reynolds~\cite{Reynolds} introduced three rules that govern the flocking behavior of agents, known as \textit{boids}, in interaction with neighboring agents. These rules, which are as follows, describe the flocking behavior of each agent: $i)$ collision avoidance: avoid collisions with nearby flockmates; $ii)$ velocity matching: attempt to match velocity with nearby flockmates; $iii)$ flock centering: attempt to stay close to nearby flockmates. The original form of these rules is that of verbal expressions, which are general and susceptible to various interpretations. A more mathematical interpretation of these rules is as follows: $i)$ separation: avoid crowding neighbors (within the repulsion zone); $ii)$ alignment: steer towards the average velocities of neighbors (within the attraction zone); $iii)$ cohesion: steer towards the average position of neighbors (within the attraction zone). Flocking studies typically divide the reaction zone of an agent within its perception range into two distinct regions: the zone of repulsion and the zone of attraction~\cite{LI2024109110,10113231}. \begin{figure}[t]
   \centering
   \includegraphics[width=0.25\textwidth]{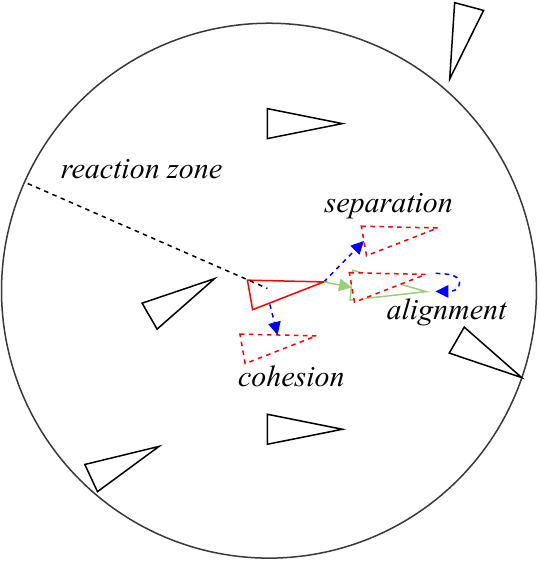}
   \includegraphics[width=0.25\textwidth]{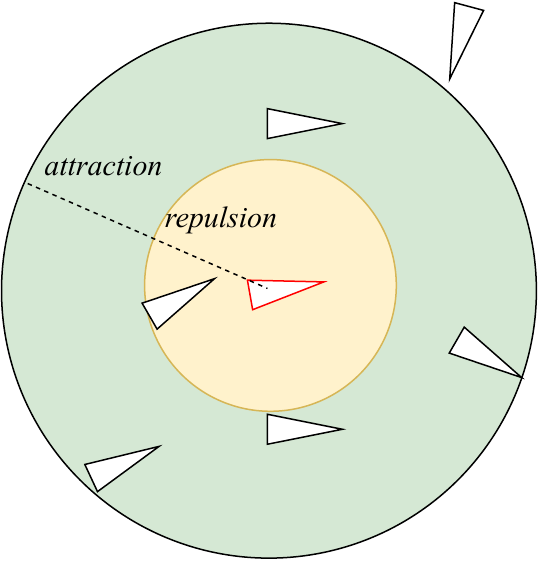}
   \includegraphics[width=0.25\textwidth]{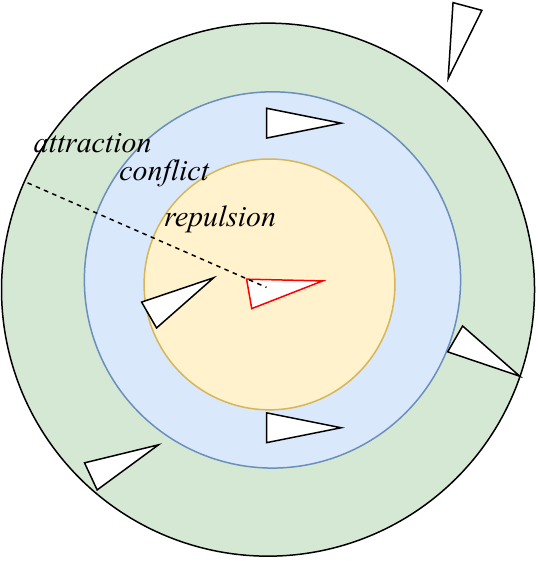}
\caption{(Left) Flocking behavior based on Reynolds \textit{boids} model. The current pose of a nominal \textit{boid} is indicated in solid red, while its future pose, determined by the Reynolds flocking rules, is shown in green. (Middle) Commonly used repulsion and attraction zones model in the literature. (Right) Proposed repulsion, conflict (repulsion and attraction), and attraction zones model. }
\label{fig:basic}
\end{figure}
A three-zone reaction model, which divides the agents' perception range into non-overlapping zones for repulsion, orientation, and attraction, has also been explored~\cite{COUZIN20021}. In these models, agents interact with each other based on the behavioral zone in which their neighbors are located. We introduce a three-zone model to effectively implement the \textit{boids} behavioral rules with the previously mentioned semi-mathematical interpretation. These zones consist of the zone of repulsion, the zone of conflict (repulsion and attraction), and the zone of attraction. Fig.~\ref{fig:basic} depicts the behavioral rules in \textit{boids} within the reaction zone, the conventional two-zone model, and our three-zone model.

Two-zone models typically use an attractive-repulsive potential function with a local minimum at a desired separation distance. The potential function exhibits a repulsive force within the repulsion zone and an attractive force within the attraction zone, depending on the pairwise relative distances of neighboring agents~\cite{LI2024109110}. Although the potential field manages both flock aggregation and collision avoidance, it has limitations, including susceptibility to local minima, scalability issues as the number of agents increases, and rigid flock formation.

Olfati-Saber~\cite{Olfati-Saber} sought to express these behavioral rules mathematically, although not by their direct mathematical translations. Three types of agents are introduced: \(\alpha\)-agents (cooperative agents), \(\beta\)-agents (obstacles), and \(\gamma\)-agents (leaders). Pairwise potential-based control laws are designed for \(\alpha\)-agent interactions to form \(\alpha\)-lattices, typically hexagonal, and to manage interactions with obstacles and leaders. The formation of the \(\alpha\)-lattice flock has gained significant attention~\cite{9998071,10644765,7112591}; however, it lacks a key property of the \textit{boids} flocking behavior, a flexible and adaptable spatial configuration. The \textit{boids} behavioral rules do not dictate a rigid lattice-type spatial formation.  Vicsek~\cite{Vicsek} and Cucker-Smale~\cite{Cucker-Smale} models have been widely acclaimed for showing cohesive flocking and consensus formation. Nevertheless, both models originally lacked flock centering, inter-agent separation, or obstacle avoidance. Moreover, these models led to rigid and occasionally regular patterns, lacking the flexibility of the \textit{boids} model.

Most existing results aimed at flocking and formation control are based on the assumption that agents exchange their position and velocity information via a high-rate communication channel, typically represented by directed or undirected graphs~\cite{aat3536,10388234,10384683}. Although this may be excessive for current multi-agent and multi-robot systems, obstacles and delays impact communication quality~\cite{9635944}. Furthermore, the limitations and unreliability of global positioning systems drive the need for alternative flocking approaches that do not depend on communication and instead utilize only the agents' onboard capabilities. In addition, the original flocking behavioral rules designed for perception-based \textit{boids} can be replicated more closely.

This paper introduces a novel flocking control approach based on local perception with zone-based interactions. Inspired by the Reynolds rules for realistic flocking behavior, each agent's perception range is partitioned into the zones of repulsion, conflict, attraction, and surveillance. By introducing the zone of conflict, where both repulsion and attraction coexist, fluid-like motion is enabled. The outer zone of surveillance is to detect and avoid alien agents. The interaction rules define the behavior of each agent, from interactions with nearby flockmates to avoidance of alien agents, obstacles, or boundaries, and adaptation to the global flock speed. The interplay of repulsive and attractive forces created by these rules produces the collective motion characteristic of flocking behavior in the real world. The desired flocking behavior for each agent is quantified within these zones with simple and perception-based expressions relying solely on bearing and distance measurements. The agent's control input is directly proportional to the quantified contribution vectors of the desired behaviors. This distributed control strategy allows MAS to achieve flexible and adaptive flocking behavior.

The proposed flocking model overcomes the limitations of traditional approaches by integrating cohesion, separation, alignment, obstacle avoidance, and alien agent avoidance into a unified framework. Unlike the Olfati-Saber model~\cite{Olfati-Saber}, which relies on indirect cohesion-separation through potential functions, our model enforces these behaviors directly while employing velocity averaging for more flexible alignment. Compared to the Vicsek model~\cite{Vicsek} and Cucker-Smale model~\cite{Cucker-Smale}, our approach accommodates variable speeds and balances all three emergent flocking behaviors: cohesion, alignment, and separation, offering a more comprehensive and adaptable solution for dynamic flocking scenarios. This unified framework positions our model as an effective and adaptable approach for simulating and controlling flocking behavior across a variety of scenarios, offering a significant improvement over traditional models.

The collective behavior of flocking exhibits emerging intelligent properties, including self-organization, robustness, adaptability, and expansibility. These properties have inspired the development of autonomous unmanned swarm systems~\cite{nwad040}. Aerial robotics swarms represent a particularly prominent area of research~\cite{MARQUEZVEGA2021100733,Soria2021,10412612,BAHAIDARAH2024101491,ZHOU2024122694}. In~\cite{MARQUEZVEGA2021100733}, a quadrotor swarm is presented to flock toward an unknown target and is optimized using multiobjective optimization. In~\cite{Soria2021}, a predictive model integrates local potential field principles into an objective function, optimizing swarm speed, order, and safety while remaining scalable and environment-independent. In~\cite{10412612}, a genetic algorithm optimizes the parameters of the adapted decentralized flocking guidance from~\cite{aat3536}, with a nonlinear sliding mode control for velocity tracking. In~\cite{BAHAIDARAH2024101491}, the Optimized Collective Motion (OCM) algorithm, utilizing viscoelastic interactions, is tuned using Particle Swarm Optimization for robust and efficient real-world applications. In~\cite{ZHOU2024122694} a dynamic flocking control framework is proposed that combines a self-propelled flocking model with a dynamic particle swarm optimization (DPSO) algorithm for adaptive control parameter optimization. 

In summary, the main contributions of this paper are as follows:  
\begin{itemize}
    \item The proposed flocking framework relies solely on bearing-distance measurements, whereas most existing models also depend on velocity measurements. 
    \item It allows individualized behavior management through weights assigned to neighbors and environmental hazards, a feature absent in generic agent models in the literature.
    \item It introduces an adaptive collision avoidance strategy, enabling agents to respond to obstacles based not only on distance but also on their motion direction and speed.
    \item It demonstrates asymptotic stability and convergence to a stable flocking configuration for any initial conditions under a spanning tree interaction graph.  
    \item Lastly, it incorporates natural predator phenomena by modeling alien agents.
\end{itemize}

The contributions outlined above, whether individually or collectively, are not addressed in existing studies, including~\cite{aat3536,MARQUEZVEGA2021100733,Soria2021,10412612,BAHAIDARAH2024101491,ZHOU2024122694}.  The remainder of this paper is organized as follows. Section~\ref{sec:zone} provides the zone-based flocking model, including the flocking rules, their mathematical translations, and the control strategy. Section~\ref{sec:bearing} derives the flocking control strategy using only bearing and distance measurements. Section~\ref{sec:simple} presents the stability analysis. Section~ \ref{sec:sim} illustrates the simulation results. Section~\ref{sec:con} concludes the paper and describes future work.

\section{Zone-Based Flocking Model}\label{sec:zone}

We propose a zone-based flocking control framework that integrates attraction-driven and repulsion-driven forces for a self-propelling agent model with a directionally aware obstacle avoidance mechanism. This framework enables flocking agents to navigate collision-free in constrained and dynamic environments while maintaining coordination and efficiency, without requiring prior knowledge of obstacle locations and environmental conditions.

\begin{figure}[t]
         \centering
         \vspace{10pt} % Adjust the vertical space as needed
         \includegraphics[width=1\textwidth]{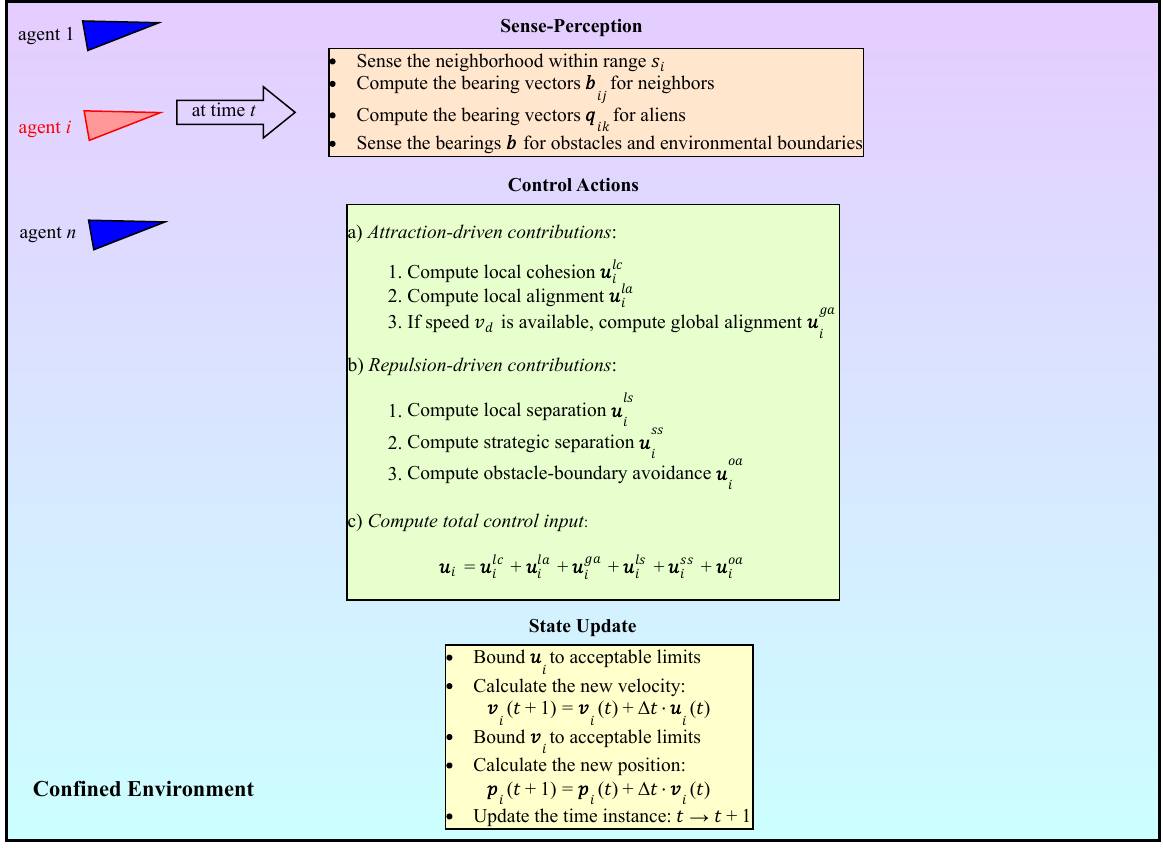}
        \caption{The architecture and flowchart of the zone-based distributed flocking control framework for interacting agents in constrained dynamic environments. The relevant quantities, definitions, and processes are detailed in the subsequent subsections.}
        \label{fig:flow}
\end{figure}

The proposed flocking controller operates locally, with each agent utilizing only local information from its neighborhood, as depicted in Fig.~\ref{fig:flow}. The framework consists of three key components, described in subsection~\ref{sec:math-exp}: (1) \textit{sense-perception}, each agent senses its neighborhood and computes bearing vectors for neighbors, aliens, obstacles, and environmental boundaries; (2) \textit{control actions}, the agent calculates attraction-driven and repulsion-driven control contributions for each sensed entity (neighbor, alien, or obstacle) and sums these contributions to determine its control input; and (3) \textit{state update}, the agent adjusts its velocity and position accordingly. This structured approach ensures that agents dynamically adapt to their surroundings while maintaining flock coherence and avoiding collisions. This proposed framework neither relies on global information nor requires the velocity of neighboring agents, making it innovative and efficient.

\subsection{Flocking Rules}\label{sec:rules}
\begin{figure}
         \centering
         \vspace{10pt} % Adjust the vertical space as needed
         \includegraphics[width=0.5\textwidth]{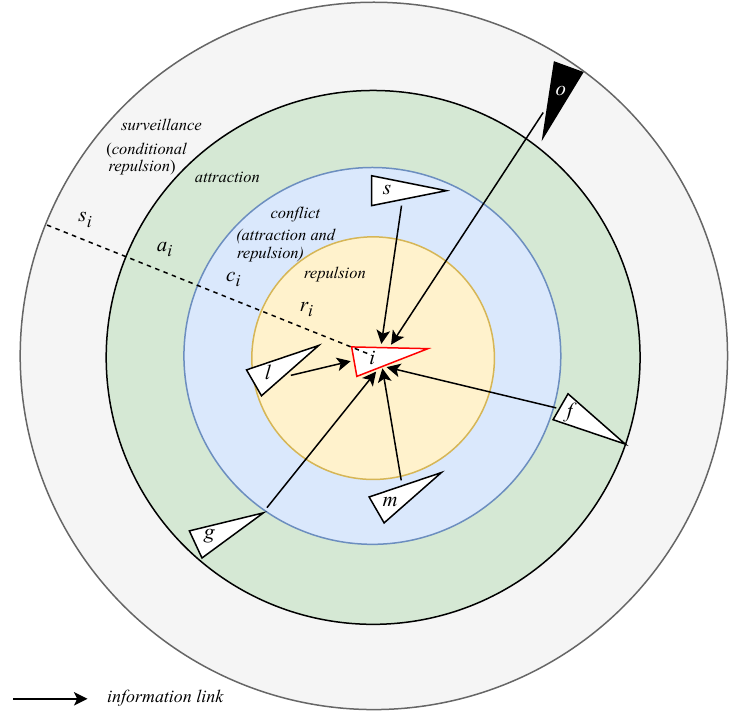}
        \caption{Repulsion, conflict (repulsion and attraction), attraction, and surveillance (conditional repulsion) zones. The set of the neighbors of agent $i$ in the repulsion, conflict, attraction, and surveillance zones, respectively, are $\mathcal{N}_i^r=\{l\}$, $\mathcal{N}_i^c=\{s,m\}$, $\mathcal{N}_i^a=\{f,g\}$, $\mathcal{N}_i^s=\{o\}$.}
        \label{fig:zones}
\end{figure}

For each individual, four concentric zones are defined: the innermost zone of repulsion, the inner annular region of conflict, the intermediate annular region of attraction, and the outermost annular region of surveillance. Without loss of generality, we consider the repulsion zone to be an interval in 1D, a circular area in 2D, and a spherical volume in 3D, all within a distance (or radius) \( r_i \). The conflict zone is the inner annular region between the repulsion zone and a larger perceptual zone, extending to a distance (or radius) \( c_i \) from the individual. The attraction zone is the intermediate annular region between the conflict zone and a further perceptual zone, extending to a distance (or radius) \( a_i \) from the individual. Finally, the surveillance zone is the outermost annular region, extending to a distance (or radius) \( s_i \) from the individual. Briefly, the reaction zone for each individual is divided into four zones as follows:
\begin{enumerate}
    \item Repulsion: the innermost zone,
    \item Conflict: the inner annular region,
    \item Attraction: the intermediate annular region, 
    \item Surveillance: the outermost annular region. 
\end{enumerate}
These zones are illustrated in Fig.~\ref{fig:zones}.

In our model, group flocking emerges from the following individual behavioral rules or interactions, based on local perception in the zones defined above. These interactions, classified into two distinct groups, are as follows:
\begin{enumerate}
    \item[] \textit{Attraction-driven:} 
    
    \item Local cohesion: Move to the weighted average position of the neighbors in the conflict and attraction zones,
          
    \item Local alignment: Move to the weighted average velocities of the neighbors in the conflict and attraction zones,
            
    \item Global speed alignment: Move to the desired global flock speed,
    
    \item[] \textit{Repulsion-driven:} 
    \item Local separation: Move away from neighbors in the repulsion and conflict zones until they are positioned at the boundary of the latter zone,
    
    \item Strategic separation: Move away from alien agents until they are positioned at the boundary of the surveillance zone,

    \item Obstacle avoidance and boundary constraints: Move away from obstacles and boundaries in the repulsion and conflict zones until they are positioned at the boundary of the latter zone.
       
\end{enumerate}

In the Reynolds \textit{boids} model, separation is local, with agents maintaining a close yet safe distance from flockmates to ensure cohesion without overcrowding. In addition to local separation, strategic separation involves adopting a larger distance from potentially dangerous alien agents to ensure safety and gain a tactical edge. Local alignment shifts individual speeds toward those of their neighbors, creating a feedback loop that eventually stabilizes the global speed of the flock. When some agents alter their speeds, their neighbors adjust accordingly, resulting in a new global speed. For smooth and coordinated group motion in a flock, it is essential to maintain a cohesive global speed tailored to the flock's specific purpose, such as foraging or escaping predators. We model natural predator phenomena as alien agents that invariably cause collisions, fragmentation, and the dispersal of flock agents. These interaction rules define each agent's behavior, from interactions with nearby flockmates to avoidance of threats, obstacles, or boundaries, and adaption with the global flock speed. The interplay of repulsive and attractive forces created by these rules produces the collective motion characteristic of flocking behavior in the real world.

\subsection{Mathematical Expressions of Flocking Rules}\label{sec:math-exp}
Consider a MAS of $n$ agents indexed by $\mathcal{V}=\{1,\cdots,n\}$. Each agent is described by the double-integrator dynamics~\cite{beaver_optimal_2020}
\begin{equation} \label{eq:double-dynamics}
     \begin{cases} 
     \dot{\mathbf{p}}_i =  \mathbf{v}_i , \\
     \dot{\mathbf{v}}_i = \mathbf{u}_i ,
   \end{cases}
 \end{equation}
where $\mathbf{p}_i,\mathbf{v}_i,\mathbf{u}_i\in\mathbb{R}^{d}$ ($d=2,3$) denote the position, velocity, and the control input of the $i$-th agent, respectively, with $i\in\mathcal{V}$. The control input for the agent \(i\) will be designed from the contributions of the flocking interactions. 

At any instant, the local control input for the agent \(i\in\mathcal{V}\) is defined as,
\begin{equation}\label{eq:control_input}
  \mathbf{u}_i =\mathbf{u}_i^{lc} + \mathbf{u}_i^{la}  + \mathbf{u}_i^{ga} + \mathbf{u}_i^{ls} + \mathbf{u}_i^{ss} + \mathbf{u}_i^{oa},  
\end{equation}
where \(\mathbf{u}_i^{lc}\), \(\mathbf{u}_i^{la}\), and \(\mathbf{u}_i^{ga}\), respectively, represent the attraction-driven contributions derived from the rules of local cohesion, local alignment, and global speed alignment interactions, \(\mathbf{u}_i^{ls}\), \(\mathbf{u}_i^{ss}\), and \(\mathbf{u}_i^{oa}\) denote the repulsion-driven contributions derived from the local separation, strategic separation, and obstacle avoidance and boundary constraints, respectively. These contribution vectors will be mathematically defined in the following.

The set of neighbors of agent $i\in\mathcal{V}$ in its repulsion, conflict, attraction, and surveillance zones are, respectively, defined as $\mathcal{N}_i^r=\{j|j\neq i\text{ and }\lVert \mathbf{p}_j-\mathbf{p}_i\rVert\leq r_i\}$, $\mathcal{N}_i^c=\{j|j\neq i\text{ and }r_i<\lVert \mathbf{p}_j-\mathbf{p}_i\rVert\leq c_i\}$, $\mathcal{N}_i^a=\{j|j\neq i\text{ and }c_i<\lVert \mathbf{p}_j-\mathbf{p}_i\rVert\leq a_i\}$, $\mathcal{N}_i^s=\{k|\text{ such that }a_i<\lVert \mathbf{q}_k-\mathbf{p}_i\rVert\leq s_i\}$, where $\mathbf{q}_k$ denotes the position of \(k\)-th alien agent and $\lVert .\rVert$ is the Euclidean norm. Note that $\mathcal{N}_i^r\cap\mathcal{N}_i^c\cap\mathcal{N}_i^a\cap\mathcal{N}_i^s=\emptyset$. For the sake of brevity, we define $\bar{\mathcal{N}}_i^r=\mathcal{N}_i^r\cup\mathcal{N}_i^c$ and $\bar{\mathcal{N}}_i^a=\mathcal{N}_i^c\cup\mathcal{N}_i^a$.

\begin{figure}
         \centering
         \vspace{10pt} % Adjust the vertical space as needed
         \includegraphics[width=0.4\textwidth]{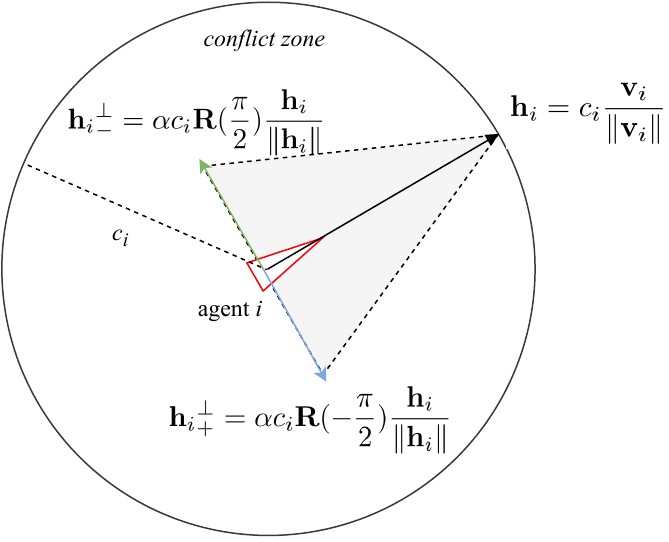}
        \caption{Triangular obstacle avoidance zone aligned with the agent's heading.}
        \label{fig:heading}
\end{figure}

For directionally aware and flexible collision avoidance with obstacles and boundary constraints, a triangular obstacle avoidance reaction zone is defined around each agent, enclosed within its conflict zone, as illustrated in Fig.~\ref{fig:heading}. The longest altitude of the triangle is aligned with the agent's heading. Thus, the obstacle avoidance reaction zone is larger in the direction of motion to accommodate the agent's need for more space to react to obstacles ahead due to its velocity. The opposite side to the longest altitude narrows the reaction zone on the agent's sides, where collisions with stationary obstacles are less likely. The configuration of the triangle (or tetrahedron in 3D) obstacle avoidance zone can adapt to the agent's speed or mission. At higher speeds, the longest altitude may extend (e.g. to the agent's attraction zone) for more reaction space, while at slower speeds, the zone may contract (e.g., to the agent's repulsion zone).

For each agent, the heading vector is defined as
\[
\mathbf{h}_i=c_i\frac{\mathbf{v}_i}{\lVert\mathbf{v}_i\rVert},
\]
where \(\mathbf{h}_i\) has a magnitude of \(c_i\). For 2D environments, two perpendicular vectors to \(\mathbf{h}_i\), formed by rotating \(\mathbf{h}_i\) by \(\pm\frac{\pi}{2}\), are 
\[
\mathbf{h}_i{^\perp_+} =\alpha c_i\mathbf{R}(-\frac{\pi}{2}) \frac{\mathbf{h}_i}{\lVert\mathbf{h}_i\rVert},\quad
\mathbf{h}_i{^\perp_-} =\alpha c_i \mathbf{R}(\frac{\pi}{2})\frac{\mathbf{h}_i}{\lVert\mathbf{h}_i\rVert},
\]
each with the magnitude of $\alpha c_i$ where $0<\alpha\leq 1$ is a scaling factor and \(\mathbf{R}\) is the rotation matrix
\[
\mathbf{R}(\theta)=\begin{bmatrix}
    \cos{(\theta)} & -\sin{(\theta)} \\ \sin{(\theta)} & \cos{(\theta)}
\end{bmatrix}.
\]
The heading vector \(\mathbf{h}_i\) represents the longest altitude of the triangular reaction zone. The vectors \(\mathbf{h}_i^{\perp_+}\) and \(\mathbf{h}_i^{\perp_-}\) correspond to the halves of the opposite side of the longest altitude of this zone in 2D. For 3D environments, these halves extend to the vertical dimension, this is crucial for avoiding objects in all directions.

The mathematical translations of the flocking interactions are as follows:
\begin{enumerate}
\item[] \textit{Attraction-driven:} 
 
\item Local cohesion:
\begin{align}
    &\mathbf{u}_i^{lc}=\frac{1}{ \sum_{j\in\bar{\mathcal{N}}_i^a}\omega_{ij}^{lc}}\sum_{j\in\bar{\mathcal{N}}_i^a}\omega_{ij}^{lc}\mathbf{p}_j-\mathbf{p}_i, \label{eq:l-cohesion}
\end{align}
where $\omega_{ij}^{lc}>0$ is the local cohesion weight that the agent $i$ assigns to each of its neighbors in $\bar{\mathcal{N}}_i^a$. The vector $\mathbf{u}_i^{lc}$ denotes agent \(i\)'s vector contributions from the desired local cohesion behavior and creates an attractive potential for that agent. 

\item Local alignment:
\begin{align}
    &\mathbf{u}_i^{la}=\frac{1}{ \sum_{j\in\bar{\mathcal{N}}_i^a}\omega_{ij}^{la}}\sum_{j\in\bar{\mathcal{N}}_i^a}\omega_{ij}^{la}\mathbf{v}_j-\mathbf{v}_i,\label{eq:l-alignment}
\end{align}
where $\omega_{ij}^{la}>0$ is the local alignment weight that the agent $i$ assigns to each of its neighbors in $\bar{\mathcal{N}}_i^a$. The vector $\mathbf{u}_i^{la}$ denotes agent \(i\)'s contributions from the desired local alignment behavior and creates an attractive potential.

\item Global speed alignment:
\begin{equation}\label{eq:g-alignment}
    \mathbf{u}_i^{ga}=\omega_i^{ga}(v^d\frac{\mathbf{v}_i}{\lVert\mathbf{v}_i \rVert}-\mathbf{v}_i),
\end{equation}
where \(v_d\) denotes the desired global speed and \(\omega_i^{ga}\geq 0\) is the global speed alignment weight or gain. If agent \(i\) is informed about \(v_d\), then \(\omega_i^{ga}>0\), otherwise \(\omega_i^{ga}=0\). For each informed agent \(i\), the vector $\mathbf{u}_i^{ga}$ denotes its contributions from the desired global speed alignment behavior and creates an attractive potential.

  \item[] \textit{Repulsion-driven:} 
  
\item Local separation:
\begin{align}\label{eq:l-separation}
    \mathbf{u}_i^{ls}=&\sum_{j\in\bar{\mathcal{N}}_i^r}\omega_{ij}^{ls}(\mathbf{p}_j-\mathbf{p}_i-c_i\frac{\mathbf{p}_j-\mathbf{p}_i}{\lVert\mathbf{p}_j-\mathbf{p}_i \rVert}), 
\end{align}
 where $\omega_{ij}^{lr}>0$ is the local separation weight that the agent $i$ assigns to each of its neighbors in $\bar{\mathcal{N}}_i^r$. The bearing vector \(\frac{\mathbf{p}_j-\mathbf{p}_i}{\lVert\mathbf{p}_j-\mathbf{p}_i \rVert}\) between agents \(i\) and \(j\) carries only directional information. When scaled by a desired distance \(c_i\), it represents the safe or comfortable local separation offset vector for agent $i$ relative to neighboring agent $j$ within agent \(i\)'s repulsion and conflict zones. The vector $\mathbf{u}_i^{ls}\in\mathbb{R}^d$ denotes agent \(i\)'s contributions from the desired local separation behavior and creates a repulsive potential for that agent. The repulsion force activates when a neighboring agent \( j \) enters the conflict zone of agent \( i \) (\(\lVert\mathbf{p}_j-\mathbf{p}_i \rVert < c_i\)). This force drives \(\lVert\mathbf{p}_j-\mathbf{p}_i \rVert\) toward \( c_i \), where it vanishes, ensuring that agent \( i \) maintains a minimum separation distance \( c_i \) from agent \( j \).

\item Strategic separation: 
\begin{align}\label{eq:s-separation}
    \mathbf{u}_i^{ss}&=\sum_{k\in\mathcal{N}_i^s}\omega_{ik}^{ss}(\mathbf{q}_k-\mathbf{p}_i-s_i\frac{\mathbf{q}_k-\mathbf{p}_i}{\lVert\mathbf{q}_k-\mathbf{p}_i \rVert}), 
\end{align}
 where $\omega_{ik}^{ss}>0$ is the strategic separation weight that the agent $i$ assigns to each alien agent (i.e., non-flockmate agent) in $\mathcal{N}_i^s$. The strategic separation offset vector $s_i\frac{\mathbf{q}_j-\mathbf{p}_i}{\lVert\mathbf{q}_j-\mathbf{p}_i \rVert}$ represents a safe directional distance from a potentially dangerous alien \(k\). The vector $\mathbf{u}_i^{ss}$ denotes agent \(i\)'s contributions from the desired strategic separation behavior and creates a repulsive potential.

\item Obstacle avoidance and boundary constraints:
\begin{align}\label{eq:o-avoidance}
    \mathbf{u}_i^{oa}&=\omega_{i}^{oa}(\mathbf{b}-\mathbf{p}_i-c_i\frac{\mathbf{b}-\mathbf{p}_i}{\lVert\mathbf{b}-\mathbf{p}_i \rVert}), 
\end{align}
 where $\omega_{i}^{oa}>0$ is the obstacle avoidance weight that the agent $i$ assigns to the obstacle within its obstacle avoidance reaction zone and \(\mathbf{b}\) denotes the closest point on the obstacle's edge or boundary to the agent. The obstacle avoidance offset vector $c_i\frac{\mathbf{b}-\mathbf{p}_i}{\lVert\mathbf{b}-\mathbf{p}_i \rVert}$ represents a safe directional distance from obstacles and boundaries. The vector $\mathbf{u}_i^{oa}$ denotes agent \(i\)'s contributions from the desired obstacle avoidance behavior and creates a repulsive potential.

Each agent continuously monitors whether its obstacle avoidance reaction zone intersects with an obstacle or boundary. A straightforward approach is to check if its heading vector \(\mathbf{h}_i\) and the perpendicular vectors (i.e., in 2D, the vectors \(\mathbf{h}_i{^\perp_+}\) and \(\mathbf{h}_i{^\perp_-}\)) intersect with an obstacle or boundary. If an intersection is detected, the agent identifies \(\mathbf{b}\). The offset vector \(\mathbf{u}_i^{oa}\) is then computed proportionally to the distance from the boundary and is used to adjust the agent's position accordingly. The triangular zone-based directional obstacle avoidance strategy focuses on obstacles directly ahead in the agent's trajectory. By prioritizing avoidance in the heading direction, the agent minimizes unnecessary reactions to obstacles that are less likely to impact its path.

\end{enumerate}

\begin{remark}
    If the agent's conflict and surveillance zones are not defined, $c_i$ and $s_i$ default to $r_i$ in \eqref{eq:l-separation},\eqref{eq:s-separation}, and \eqref{eq:o-avoidance}.
\end{remark}

%\begin{remark}    Given the magnitude of the position vectors, it is imperative to scale down the magnitude of \(\mathbf{e}_{ij}^{lc}\) while preserving its direction to ensure consistency. One straightforward method is to normalize \(\mathbf{u}_i^{lc}\), and then multiply it by a positive weight.\end{remark}

%\subsection{Flocking Controller}

%The contribution of an agent from the desired flocking behavior, as defined in \eqref{eq:l-separation}–\eqref{eq:g-alignment}, represents the flocking requirements for that agent. The total contribution vector for agent \(i \in \mathcal{V}\) from the desired flocking behavior is expressed as \begin{equation}\label{eq:deviation}  \mathbf{e}_i = \mathbf{e}_i^{ls} + \mathbf{e}_i^{lc} + \mathbf{e}_i^{la} + \mathbf{e}_i^{ss} + \mathbf{e}_i^{oa} + \mathbf{e}_i^{ga}.   \end{equation}

%Let $\mathbf{u}_i\in\mathit{\mathbb{R}}^d$ denote the control input vector for agent \(i\in\mathcal{V}\). We define \begin{equation}\label{eq:control}    \mathbf{u}_i=g_i\mathbf{e}_i,\end{equation} where $g_i$ is the controller gain. 

\begin{remark}
With boundary conditions, the environment is confined, and therefore, even if some agents become separated from the flock or the flock fragments, the agents will eventually reunite. Without boundary conditions, it is necessary to assume that each agent always has at least one flockmate within its conflict or attraction zone. To ensure this, a straightforward approach is to gradually increase the radius of the attraction zone until a flockmate is detected.
\end{remark}

Unifying each agent's repulsion and attraction potentials into the control input~\eqref{eq:control_input} creates an artificial flocking potential energy for that agent. The control input of agent \(i\) are bounded as follows,
\begin{align}
   & \mathbf{u}_i = u_i^{\max} \tanh{(\frac{\lVert \mathbf{u}_i \rVert}{u_i^{\max}})} \frac{\mathbf{u}_i}{\lVert \mathbf{u}_i \rVert}, \label{eq:u-bound}
\end{align}
where \( u_i^{\max} \) are the maximum acceleration magnitudes for agent \(i\). These smooth cutoffs at \( u_i^{\max} \) maintain the direction while reducing the magnitude.

The resultant velocity from \eqref{eq:u-bound} for physical agents is subject to saturation. To prevent velocity saturation, the velocity of agent \(i\) is restricted as follows,
\begin{align}
   & \mathbf{v}_i = v_i^{\max} \tanh{(\frac{\lVert \mathbf{v}_i \rVert}{v_i^{\max}})} \frac{\mathbf{v}_i}{\lVert \mathbf{v}_i \rVert}, \label{eq:v-bound} 
\end{align}
where \( v_i^{\max} \) is the maximum velocity magnitudes for agent \(i\). If the computed velocity exceeds the agent's limits, it is smoothly capped at \( v_i^{\max} \).

\section{Flocking with Bearing-Distance Measurements} \label{sec:bearing}

In general, MAS control can be classified into position-based, displacement-based, and distance-based control~\cite{OH2015424}. The controller in \eqref{eq:control_input} falls under the position-based category. In this approach, agents sense their positions relative to a global coordinate system and actively adjust their positions to meet the desired behavior, defined in terms of desired positions within the global coordinate system. Consequently, this method requires agents to be equipped with precise global positioning sensors. Let \(\mathbf{p}_{ij} = \mathbf{p}_j - \mathbf{p}_i\) be the relative position vector , with \(\dot{\mathbf{p}}_{ij} = \mathbf{v}_j - \mathbf{v}_i\). If the relative positions \(\mathbf{p}_{ij}\) are sensed with respect to the global coordinate system, meaning the agents know the orientation of the global coordinate system, all flocking behavioral contribution vectors can be re-expressed in terms of \(\mathbf{p}_{ij}\) and \(\dot{\mathbf{p}}_{ij}\). As a result, the controller in \eqref{eq:control_input} transitions to a displacement-based control approach, where agents no longer need to sense their absolute positions. In distance-based control, agents actively regulate the distances between themselves and their neighbors to achieve the desired behavior defined by the inter-agent distances. Each agent senses the relative positions of its neighbors with respect to its own local coordinate system, with the orientations of these local coordinate systems not necessarily being aligned. While distance-based control offers advantages in terms of reduced sensing requirements, it demands more interactions among agents to maintain the desired behavior. Since flocking agents, like \textit{boids}, are perception-driven interactive entities, displacement-based and distance-based control approaches are more practical for flocking. In the following, we derive the desired behavioral contribution vectors for flocking based on bearing measurements and inter-agent distances. This eliminates the need for agents to have knowledge of the global coordinate system or their positions.

Let the bearing vector \(\mathbf{b}_{ij}=\frac{\mathbf{p}_j-\mathbf{p}_i}{\lVert\mathbf{p}_j-\mathbf{p}_i \rVert}\) and inter-agent distance \(d_{ij}=\lVert\mathbf{p}_j-\mathbf{p}_i \rVert\) be available from the onboard sensors. Their rate of change is also either measured directly or computed from the sensor data. The bearing vector \(\mathbf{b}_{ij}\) and inter-agent distance \(d_{ij}\) are related by
\begin{equation}\label{eq:related}
\mathbf{b}_{ij}=\frac{\mathbf{p}_j-\mathbf{p}_i}{d_{ij}}.
\end{equation}

The time derivative of the bearing vector \(\mathbf{b}_{ij}\) is given by
\begin{equation}\label{eq:deriv_bij}
   \dot{\mathbf{b}}_{ij} = \frac{1}{d_{ij}}( (\mathbf{v}_j - \mathbf{v}_i)- \dot{d}_{ij}\mathbf{b}_{ij}),
\end{equation}
or,
\begin{equation}\label{eq:bear-vel}
   \mathbf{v}_j - \mathbf{v}_i = d_{ij}\dot{\mathbf{b}}_{ij}+\dot{d}_{ij}\mathbf{b}_{ij}. 
\end{equation}

The interaction vectors for flocking, based on bearing measurements and inter-agent distances, are derived as follows:
\begin{enumerate}
\item[] \textit{Attraction-driven:} 

\item Bearing-distance local cohesion:
Multiplying both sides of local cohesion in \eqref{eq:l-cohesion} by $\sum_{j\in\bar{\mathcal{N}}_i^{la}}\omega_{ij}^{lc}$ yields 
\begin{align*}
    \sum_{j\in\bar{\mathcal{N}}_i^a}\omega_{ij}^{lc}\mathbf{u}_i^{lc}&=\sum_{j\in\bar{\mathcal{N}}_i^a}\omega_{ij}^{lc}\mathbf{p}_j-\sum_{j\in\bar{\mathcal{N}}_i^a}\omega_{ij}^{lc}\mathbf{p}_i\notag\\&=\sum_{j\in\bar{\mathcal{N}}_i^a}\omega_{ij}^{lc}(\mathbf{p}_j-\mathbf{p}_i)=\sum_{j\in\bar{\mathcal{N}}_i^a}\omega_{ij}^{lc}d_{ij}\mathbf{b}_{ij},
\end{align*}
or,
\begin{align}\label{eq:b-l-cohesion}
    \mathbf{u}_i^{lc}=\frac{1}{\sum_{j\in\bar{\mathcal{N}}_i^a}\omega_{ij}^{lc}}\sum_{j\in\bar{\mathcal{N}}_i^{la}}\omega_{ij}^{lc}d_{ij}\mathbf{b}_{ij},
\end{align}
which is the bearing-distance local cohesion contribution vector for the agent \(i\).   

\item Bearing-distance local alignment: Similarly to the bearing-distance local cohesion, we arrive at
\[
\mathbf{u}_i^{la}=\frac{1}{\sum_{j\in\bar{\mathcal{N}}_i^a}\omega_{ij}^{la}}\sum_{j\in\bar{\mathcal{N}}_i^a}\omega_{ij}^{la}(\mathbf{v}_j-\mathbf{v}_i).
\]
Using \eqref{eq:bear-vel}, the bearing-distance local alignment contribution vector for agent \(i\) is given by
\begin{align}\label{eq:b-l-alignment}
    \mathbf{u}_i^{la}=\frac{1}{\sum_{j\in\bar{\mathcal{N}}_i^a}\omega_{ij}^{la}}\sum_{j\in\bar{\mathcal{N}}_i^a}\omega_{ij}^{la}(d_{ij}\dot{\mathbf{b}}_{ij}+\dot{d}_{ij}\mathbf{b}_{ij}).
\end{align}

\item Bearing-distance global speed alignment: Define the auxiliary vector
\begin{equation}\label{eq:aux}
\mathbf{u}_i^{\text{aux}}=\mathbf{u}_i^{ls}+\mathbf{u}_i^{lc}+\mathbf{u}_i^{la}+\mathbf{u}_i^{ss}+\mathbf{u}_i^{oa}.
\end{equation}
For an informed agent \(i\) with the desired global speed \(v^d\), its velocity is estimated based on the bearing-distance measurements as follows,
\begin{equation}\label{eq:vel-est}
\hat{\mathbf{v}}_i= \hat{\mathbf{v}}_i^{\text{prev}} +\delta\mathbf{u}_i^{\text{aux}},
\end{equation}
where \(\hat{\mathbf{v}}_i^{\text{prev}}\) denotes the previously estimated velocity at the prior time step, and \(\delta\) is the time step duration. Accordingly, the bearing-distance global speed alignment contribution vector for agent \(i\) is given by
\begin{equation}\label{eq:b-g-alignment}
    \mathbf{u}_i^{ga}=\omega_i^{ga}(v^d\frac{\hat{\mathbf{v}}_i}{\lVert\hat{\mathbf{v}}_i \rVert}-\hat{\mathbf{v}}_i).
\end{equation}

\item[] \textit{Repulsion-driven:} 
 
\item Bearing-distance local separation:
By substituting the bearing vector \(\mathbf{b}_{ij}\) into \eqref{eq:l-separation} and using \eqref{eq:related}, we obtain the bearing-distance  local separation contribution vector for the agent \(i\) as
\begin{align}\label{eq:b-l-separation}
    \mathbf{u}_i^{ls}=&\sum_{j\in\bar{\mathcal{N}}_i^r}\omega_{ij}^{ls}(d_{ij}-c_i)\mathbf{b}_{ij}. 
\end{align}

\item Bearing-distance strategic separation: 
Denote \(\mathbf{b}_{ik}\) as the bearing vector between agent \(i\) and an alien agent \(k\), and \(d_{ik}\) as the distance between them. The bearing-distance strategic separation contribution vector for agent \(i\) is given by
\begin{align}\label{eq:b-s-separation}
    \mathbf{u}_i^{ss}=&\sum_{j\in\mathcal{N}_i^s}\omega_{ik}^{ss}(d_{ik}-s_i)\mathbf{b}_{ik}. 
\end{align}

\item Bearing-distance obstacle avoidance and boundary constraints:
Denote \(\mathbf{b}_{ie}\) as the bearing vector between agent \(i\) and a boundary edge (of an obstacle or environment), and \(d_{ie}\) as the distance between them. The bearing-distance obstacle avoidance and boundary constraints contribution vector for agent \(i\) is given by
\begin{align}\label{eq:b-o-avoidance}
    \mathbf{u}_i^{oa}&=\sum\omega_{i}^{oa}(d_{ie}-c_i)\mathbf{b}_{ie}.
\end{align}

\end{enumerate}

Given that all interaction vectors are derived in terms of bearing-distance measurements, the bearing-distance flocking control input vector for agent \(i\in\mathcal{V}\) is computed as
\begin{equation}\label{eq:b-control}
    \mathbf{u}_i=\mathbf{u}_i^{\text{aux}}+\mathbf{u}_i^{ga}.
\end{equation}

The proposed distributed flocking scheme is summarized in Algorithm~\ref{algo}. For each agent, the interaction vectors for local separation, local cohesion, local alignment, strategic separation, and obstacle avoidance using bearing vectors and inter-agent distances. These interaction vectors are aggregated to form an auxiliary vector to estimate the agent's velocity in alignment with a desired global speed. The control input for each agent is then computed by combining the auxiliary vector with the global speed alignment vector. This distributed approach enables adaptive flocking control based on local interactions. In the algorithm, \(m\) denotes the number of alien agents, \(o\) represents the number of obstacles, and \(\mathbf{p}^{\min}\) and \(\mathbf{p}^{\max}\) specify the boundary conditions. 

\begin{remark}
 We derived the bearing-distance controller from the position-based controller given in the previous section, rendering them equivalent. Therefore, in scenarios where bearing-distance measurements are unavailable but position data is accessible, the position-based controller can achieve flocking behavior consistent with that of the bearing-distance-based controller.   
\end{remark}

\begin{algorithm}[h]
\caption{Bearing-Distance Flocking with Zone-Based Interactions}
\begin{algorithmic}[1] \label{algo}

    \STATE Set \( d \), \( n \), \(v_d\),  \(m\), \(o\), \(\mathbf{p}^{\min},\mathbf{p}^{\max}\).
     \STATE Initialize  \(o\) obstacles.
    \STATE Initialize \( \mathbf{p}_i\) for \( i\in\{1,\cdots,n\} \) and \(\mathbf{p}_k^a\) for \( k\in\{1,\cdots,m\} \).
    \STATE Initialize \( r_i \), \( c_i \), \( a_i \), \( s_i \), \( g_i \), \(v_i^{\max}\), \( \mathbf{v}_i\), \(u_i^{\max}\), \( \omega_{ij}^{ls} \), \( \omega_{ij}^{lc} \), \( \omega_{ij}^{la} \), \( \omega_{ik}^{ss} \), \( \omega_i^{oa} \), \( \omega_i^{ga} \) for \( i,j\in\{1,\cdots,n\} \), and \(v_k^{a\max}\), \(\mathbf{v}_k^a\) for \( k\in\{1,\cdots,m\} \).
    
    \FOR{each time step}
        \FOR{each agent \( i\in\{1,\cdots,n\} \)}
            \STATE Identify neighbors \( \mathcal{N}_i^r \), \( \mathcal{N}_i^c \), \( \mathcal{N}_i^a \), and \( \mathcal{N}_i^s \) based on distance measurements \(d_{ij}\), \(d_{ik}\),  \(d_{ie}\).
            \STATE Measure the bearing vectors \(\mathbf{b}_{ij}\), \(\mathbf{b}_{ik}\), \(\mathbf{b}_{ie}\). 
            \STATE Calculate the separation interaction \( \mathbf{u}_i^{ls}\) from \eqref{eq:b-l-separation}.
            \STATE Calculate the cohesion interaction \( \mathbf{u}_i^{lc}\) from \eqref{eq:b-l-cohesion}.
            \STATE Calculate the alignment interaction \( \mathbf{u}_i^{la}\) from \eqref{eq:b-l-alignment}.
            \STATE Calculate the strategic separation interaction \( \mathbf{u}_i^{ss}\) from \eqref{eq:b-s-separation}.  
            \STATE Calculate the obstacle avoidance and boundary conditions interaction \( \mathbf{e}_i^{oa}\) in \eqref{eq:b-o-avoidance}.   
            \STATE Calculate auxiliary \(\mathbf{u}_i^{\text{aux}}\) from \eqref{eq:aux}.
            \STATE Calculate \(\hat{\mathbf{v}}_i\) from \eqref{eq:vel-est}.
            \STATE Calculate the global speed alignment interaction \( \mathbf{u}_i^{ga}\) from \eqref{eq:b-g-alignment}.
            \STATE Calculate control input \( \mathbf{u}_i \) from \eqref{eq:b-control}.
            \STATE Bound control input \( \mathbf{u}_i \) from \eqref{eq:u-bound}.
            \STATE Update velocity \( \mathbf{v}_i \).
            \STATE Bound velocity \( \mathbf{v}_i \) from \eqref{eq:v-bound}.
            \STATE Update position \( \mathbf{p}_i \).
        \ENDFOR
        \FOR{each alien \( k \in \{1,\cdots,m\} \)}
         \STATE Update \( \mathbf{p}_k^a \).
        \ENDFOR
    \ENDFOR
\end{algorithmic}
\end{algorithm}

%\textit{Boids}-like flocking is unstructured, allowing agents to rearrange freely as long as they maintain cohesion, separation, and alignment. In contrast, lattice-type flocking enforces a regular geometric arrangement.

%\begin{remark}When \(c_i = c\forall i\in\mathcal{V}\), and incorporating the separation \eqref{eq:b-l-separation} alongside the alignment \eqref{eq:b-l-alignment} while omitting the cohesion \eqref{eq:b-l-cohesion}, the flock adopts a lattice-type geometric configuration.\end{remark}

\section{Stability Analysis}\label{sec:simple}
For the purpose of conducting a stability analysis, we introduce a minimal stand-alone flocking model derived from the zone-based flocking control framework.

Without loss of generality, consider a minimal flocking control input defined as
\begin{equation}\label{eq:simple-con}
    \mathbf{u}_i^c=\mathbf{u}_i^a+\mathbf{u}_i^s,
\end{equation}
consisting of an attraction-driven alignment contribution vector given by
  \begin{equation}\label{eq:ali}
  \mathbf{u}_i^a= \frac{1}{ \sum_{j\in\mathcal{N}_i}\sigma_{ij}}\sum_{j\in\mathcal{N}_i}\sigma_{ij}\mathbf{v}_j-\mathbf{v}_i,
    \end{equation}
and a repulsion-driven separation contribution vector given by
    \begin{equation}\label{eq:sep-app}
   \mathbf{u}_i^s=\sum_{j\in\mathcal{N}_i}\mu_{ij}(\mathbf{p}_j-\mathbf{p}_i-\alpha_i r_i\frac{\mathbf{p}_j-\mathbf{p}_i}{\lVert \mathbf{p}_j-\mathbf{p}_i\rVert}),
   \end{equation}
where \(\alpha_i=\beta_i|\mathcal{N}_i|\), \(\beta_i\) is a scaling factor, typically chosen to be inversely proportional to \(r_i\), e.g., \(\beta_i = \frac{1}{r_i}\), \(r_i\) here is the radius of the perception zone for agent \(i\), \(\mathcal{N}_i\) denotes the set of neighbors of agent \(i\) within its perception zone, \(|\mathcal{N}_i|\) denotes the cardinality of \(\mathcal{N}_i\), and \(\sigma_{ij},\mu_{ij}>0\). 

Note that, within the zone-based flocking control framework outlined in Section~\ref{sec:zone}, both terms coexist when a neighboring agent \(j\) is within agent \(i\)'s conflict region.

In the minimal flocking control model, the separation contribution vector in~\eqref{eq:sep-app} alone results in oscillatory behavior; agents continually move away from neighbors within the range \( \alpha_i r_i \) and then approach them once they are outside this range. However, this rule by itself does not help agents establish a common motion direction or stabilize a sufficient separation distance to prevent collisions. Incorporating the alignment contribution in~\eqref{eq:ali} addresses both issues, enabling agents to achieve coordinated motion and avoid close encounters.

We conduct a closed-loop system stability analysis to investigate whether the control strategies~\eqref{eq:simple-con} achieve stable MAS flocking from an arbitrary initial condition. To construct a closed-loop system for stability analysis in flocking through local feedback control strategies, we rely on some basic definitions and concepts in algebraic graph theory and an assumption on the dynamics of the agent motion, which are given in the following.

A directed graph is a pair $\mathcal{G}(\mathcal{V},\mathcal{E})$ where $\mathcal{V}$ is a finite set of vertices or nodes and $\mathcal{E}\subseteq \{(j,i):i,j \in \mathcal{V}\}$ is a set of edges or arcs. Each edge $(j,i)\in \mathcal{E}$ represents an information flow from node $j$ to node $i$. The graph has no self-loops, i.e., for all $(j,i)\in \mathcal{E},i\neq j$. The set of (incoming) neighbors of vertex $i$ is defined by $\mathcal{N}_i^{\text{in}}=\{j \in \mathcal{V} \mid j \neq i, \,(i,j)\in \mathcal{E}\}$, where it consists of all vertices \( j \) that have a directed edge pointing from \( j \) to \( i \). Node \(i\) is globally reachable in graph $ \mathcal{G}$ if there exists a sequence of edges directed from node \( j \in \mathcal{V}, j \neq i \), to node \( i \). A globally reachable node is also known as a root node of a spanning tree in the graph. Graph \( \mathcal{G} \) is connected if at least one globally reachable node exists. A directed tree is a connected directed graph where every node except one, called the root, has an in-degree equal to one. A spanning tree of a directed graph is a directed tree formed by edges of the graph that connect all the nodes. 

The directed graph definition above is well suited to represent the information flow or interactions between agents in a MAS. Each agent is represented as a node, with the (directed) edges indicating the directional perceptions between agents. If agent \(i\) perceives agent \(j\), there is a directed edge from \(j\) to \(i\), denoted as \( (j,i) \in \mathcal{E} \) and is assigned with weights \(\mu_{ij},\mu_{ij}\). In this context, \(\mathcal{N}_i^{\text{in}}\) represents the set of nodes from which node \( i \) receives information or influence. It consists of all agents \(j\) that are within agent \(i\)'s perception range, i.e., \(\mathcal{N}_i^{\text{in}}\equiv \mathcal{N}_i =\left\{ j \in \mathcal{V} \mid j \neq i, \, \lVert \mathbf{p}_j - \mathbf{p}_i \rVert \leq r_i \right\}\). 

To maintain cohesive flocking behavior within the MAS, graph connectivity is necessary. Thus, we make the following assumption.
\begin{assumption}\label{assump:connect}
    The information flow (or interaction) graph $\mathcal{G}(\mathcal{V},\mathcal{E})$ of the MAS remains connected at all times.
\end{assumption}

Matrix $\mathbf{D} \in \mathit{\mathbb{R}} ^{\mathcal{|V|}\times\mathcal{|E|}}$ is the incidence matrix of $\mathcal{G}$, where $\mathbf{D}$'s $ki$-th element is $1$ if the node $k$ is the head (receiving) node of the edge $i$, $-1$ if the node $i$ is the tail (sending) node of the edge \(i\), and $0$, otherwise. The graph Laplacian matrix is defined as \(\mathbf{L}= \mathbf{D} \mathbf{D}^\top \in \mathbb{R}^{|\mathcal{V}| \times |\mathcal{V}|}\). For a \(d\)-dimensional graph, the extended Laplacian matrix is given by \(\mathbf{L} \otimes \mathbf{I}_d = (\mathbf{D} \mathbf{D}^\top) \otimes \mathbf{I}_d \in \mathbb{R}^{d|\mathcal{V}| \times d|\mathcal{V}|}\), where \(\mathbf{I}_d\) is the identity matrix of dimension \(d\), and \(\otimes\) denotes the Kronecker product~\cite{Olfati-Saber}. Notably, a key property of the Laplacian matrix is that all its eigenvalues have nonnegative real parts.

%A primary application for the proposed flocking control strategies is group flocking of unmanned aerial vehicles (UAVs). The nonlinear dynamics of a fixed-wing UAV can be simplified to a double-integrator model using the feedback linearization technique (e.g., see the linearization for the fixed-wing UAVs in~\cite{lin2014distributed,wang2012integrated}). %For quadrotor UAVs, the dynamics are mapped into a flat space, where the quadrotor's nonlinear dynamics are simplified to a triple-integrator model (see~\cite{5980409}). 
The dynamics of a team of \(n\) agents are governed by the following system
\begin{equation} \label{eq:dynamics}
     \begin{cases} 
     \dot{\mathbf{p}} =  \mathbf{v} , \\
     \dot{\mathbf{v}} = \mathbf{u} ,
   \end{cases}
 \end{equation}
where \(\mathbf{p}=[\mathbf{p}_1^\top,\cdots,\mathbf{p}_n^\top]^\top\), \(\mathbf{v}=[\mathbf{v}_1^\top,\cdots,\mathbf{v}_n^\top]^\top\), and \(\mathbf{u}=[\mathbf{u}_1^\top,\cdots,\mathbf{u}_n^\top]^\top\). %While ~\eqref{eq:dynamics} describes the team dynamics using a double-integrator agent model, which will be employed for constructing the closed-loop system and performing its stability analysis, the processes for a team using a triple-integrator agent model are similar. 

Within the minimal flocking control framework, the flocking behavior is achieved when all agents are driven such that the edge interactions asymptotically converge to zero, i.e.,
\begin{equation} \label{eq:requirement}
\mu_{ij}(\mathbf{p}_j-\mathbf{p}_i-\mathbf{q}_{ij})\rightarrow 0,\quad\frac{1}{\sum_{j\in\mathcal{N}_i}\sigma_{ij}}\sum_{j\in\mathcal{N}_i}\sigma_{ij}\mathbf{v}_j-\mathbf{v}_i\rightarrow 0, \text{ for all }(j,i)\in\mathcal{E},\text{ as } t\rightarrow\infty,
\end{equation}
where \(\mathbf{q}_{ij}=\alpha_i r_i\frac{\mathbf{p}_j-\mathbf{p}_i}{\lVert \mathbf{p}_j-\mathbf{p}_i\rVert}\).

Let $\mathbf{q}=[\cdots,\mathbf{q}_{ij}^\top,\cdots]^\top\in\mathbb{R}^{d|\mathcal{E}|}$ where the order of the elements correspond to the order of the edges in \(\mathcal{E}\). We can rewrite the separation component in the control strategies~\eqref{eq:simple-con} as
\begin{align}
    &\sum_{j\in\mathcal{N}_i}\mu_{ij}(\mathbf{p}_j-\mathbf{p}_i-\mathbf{q}_{ij})=-(\mathbf{d}_i\mathbf{W}_i^s\mathbf{D}^\top\otimes \mathbf{I}_d) \mathbf{p}+(\mathbf{d}_i\mathbf{W}_i^s\otimes \mathbf{I}_d)\mathbf{q}, \label{eq:sum0}
\end{align}
where $\mathbf{d}_i$ denote the $i$-th row in $\mathbf{D}$ and \(\mathbf{W}_i^s = \text{diag}(0,\cdots, \mu_{ij}, \cdots,0) \) for all \(j \in \mathcal{N}_i \in \mathbb{R}^{|\mathcal{E}|}\).

The following holds
\begin{align}
    &\sum_{j\in\mathcal{N}_i}\sigma_{ij}(\mathbf{v}_j-\mathbf{v}_i)=-(\mathbf{d}_i\mathbf{W}_i^a\mathbf{D}^\top\otimes \mathbf{I}_d) \mathbf{v},\label{eq:sum1}
\end{align}
where  \(\mathbf{W}_i^a = \text{diag}(0,\cdots, \sigma_{ij}, \cdots,0)\) for all \(j \in \mathcal{N}_i \in \mathbb{R}^{|\mathcal{E}|}\). Expanding (\ref{eq:sum1}), we have
\begin{equation}\label{eq:sum2}
   \sum_{j\in\mathcal{N}_i}\sigma_{ij}\mathbf{v}_j- \mathbf{v}_i\sum_{j\in\mathcal{N}_i}\sigma_{ij}=- (\mathbf{d}_i\mathbf{W}_i^a\mathbf{D}^\top\otimes \mathbf{I}_d) \mathbf{v}.
\end{equation}
Define $\bar{\mathbf{W}}_i^a=\frac{1}{\sum_{j\in\mathcal{N}_i}\sigma_{ij}}\mathbf{W}_i^a$. Multiplying both sides of (\ref{eq:sum2}) by $\frac{1}{\sum_{j\in\mathcal{N}_i}\sigma_{ij}}$ yields the alignment component in the following form
\begin{equation}\label{eq:sum3}
    \frac{1}{\sum_{j\in\mathcal{N}_i}\sigma_{ij}}\sum_{j\in\mathcal{N}_i}\sigma_{ij}\mathbf{v}_j-\mathbf{v}_i=-(\mathbf{d}_i\bar{\mathbf{W}}_i^a\mathbf{D}^\top\otimes \mathbf{I}_d) \mathbf{v}.
\end{equation}

Using~\eqref{eq:sum0} and~\eqref{eq:sum3}, the control strategies~\eqref{eq:simple-con} in compact form are given by
\begin{align}\label{eq:control_comp_i}
    \mathbf{u}_i&=-(\mathbf{d}_i\mathbf{W}_i^s\mathbf{D}^\top\otimes \mathbf{I}_d) \mathbf{p}+(\mathbf{d}_i\mathbf{W}_i^s\otimes \mathbf{I}_d)\mathbf{q}-(\mathbf{d}_i\bar{\mathbf{W}}_i^a\mathbf{D}^\top\otimes \mathbf{I}_d) \mathbf{v}.
\end{align}
Stacking~\eqref{eq:control_comp_i} for all \(i\in\mathcal{V}\) yields
\begin{align}\label{eq:control_comp}
    \mathbf{u}&=-(\begin{bmatrix}
        \mathbf{d}_1\mathbf{W}_1^s \\ \vdots \\ \mathbf{d}_n\mathbf{W}_n^s
    \end{bmatrix}\mathbf{D}^\top\otimes \mathbf{I}_d) \mathbf{p}-(\begin{bmatrix}
        \mathbf{d}_1\bar{\mathbf{W}}_1^a \\ \vdots \\ \mathbf{d}_n\bar{\mathbf{W}}_n^a
    \end{bmatrix}\mathbf{D}^\top\otimes \mathbf{I}_d) \mathbf{v}+(\begin{bmatrix}
      \mathbf{d}_1\mathbf{W}_1^s\\ \vdots \\ \mathbf{d}_n\mathbf{W}_n^s
    \end{bmatrix}\otimes \mathbf{I}_d)\mathbf{q}.
\end{align}

Note that 
\[
(\begin{bmatrix}
      \mathbf{d}_1\mathbf{W}_1^s \\ \vdots \\ \mathbf{d}_n\mathbf{W}_n^s
    \end{bmatrix}\otimes \mathbf{I}_d)\mathbf{q}=(\begin{bmatrix}
    \alpha_1 z_1\mathbf{d}_1\mathbf{W}_1^s\mathbf{W}_1^q \\ \vdots \\ \alpha_n z_n\mathbf{d}_n\mathbf{W}_n^s\mathbf{W}_n^q
\end{bmatrix}\mathbf{D}^\top\otimes \mathbf{I}_d) \mathbf{p},
\]
where \(\mathbf{W}_i^q = \text{diag}(0,\cdots, \frac{1}{\lVert \mathbf{p}_j-\mathbf{p}_i\rVert}, \cdots,0)\) for all \(j \in \mathcal{N}_i \in \mathbb{R}^{|\mathcal{E}|}\). Thus,~\eqref{eq:control_comp} is simplified as
\begin{align}\label{eq:control_comp2}
    \mathbf{u}&=-(\begin{bmatrix}
        \mathbf{d}_1\bar{\mathbf{W}}_1^s \\ \vdots \\ \mathbf{d}_n\bar{\mathbf{W}}_n^s
    \end{bmatrix}\mathbf{D}^\top\otimes \mathbf{I}_d) \mathbf{p}-(\begin{bmatrix}
        \mathbf{d}_1\bar{\mathbf{W}}_1^a \\ \vdots \\ \mathbf{d}_n\bar{\mathbf{W}}_n^a
    \end{bmatrix}\mathbf{D}^\top\otimes \mathbf{I}_d) \mathbf{v},
\end{align}
where \(\bar{\mathbf{W}}_i^p=\mathbf{W}_i^s(\mathbf{I}-\alpha_i r_i\mathbf{W}_i^q)\).

Denote
\[
\mathbf{D}^s = \begin{bmatrix}
         \mathbf{d}_1 \bar{\mathbf{W}}_1^s \\ \vdots \\  \mathbf{d}_n \bar{\mathbf{W}}_n^s
    \end{bmatrix} \otimes \mathbf{I}_d, \quad \mathbf{D}^a = \begin{bmatrix}
         \mathbf{d}_1 \bar{\mathbf{W}}_1^a \\ \vdots \\ \mathbf{d}_n \bar{\mathbf{W}}_n^a
    \end{bmatrix} \otimes \mathbf{I}_d,
\]
where both matrices have a structure similar to the incidence matrix with elements scaled by weights embedded in the diagonals of \(\bar{\mathbf{W}}_i^p\) and \(\bar{\mathbf{W}}_i^a\) for each node \(i\). The control strategies \eqref{eq:control_comp2} in their global form are given by
\begin{equation}\label{eq:cont-global}
     \mathbf{u}=-\mathbf{D}^s(\mathbf{D}^\top\otimes \mathbf{I}_d)\mathbf{p} -\mathbf{D}^a(\mathbf{D}^\top\otimes \mathbf{I}_d) \mathbf{v}.
\end{equation}

Let \(\mathbf{x}=(-\mathbf{D}^\top\otimes \mathbf{I}_d) \mathbf{p}\) and \(\dot{\mathbf{x}}=(-\mathbf{D}^\top\otimes \mathbf{I}_d) \mathbf{v}\) where the vectors \(\mathbf{x}\) and \(\dot{\mathbf{x}}\) contain all edge interactions. Then,  \(\ddot{\mathbf{x}}=(-\mathbf{D}^\top\otimes \mathbf{I}_d) \mathbf{u}\). The flocking behavior requirement in \eqref{eq:requirement} is equivalent to 
\[
\mathbf{x}\rightarrow0, \quad\dot{\mathbf{x}}\rightarrow 0, \text{ as } t\rightarrow\infty.
\]

\begin{theorem}\label{theorem:stability}
Consider a MAS with the team dynamics in~\eqref{eq:dynamics} and the control strategies given by~\eqref{eq:cont-global}. Let the corresponding interaction graph \(\mathcal{G}\) be a spanning tree. Then, the edge interactions system will asymptotically converge to zero, i.e., 
\begin{equation}\label{eq:AS}
  \lim_{t \to \infty}  \begin{bmatrix}
        \mathbf{x} \\ \dot{\mathbf{x}}
    \end{bmatrix}\rightarrow 0,
\end{equation}
thereby ensuring the MAS flocking behavior. 
\end{theorem}
\begin{proof}
The edge interaction dynamics can be expressed as
\begin{align}
\begin{bmatrix}\label{eq:edge-dynamics}
      \dot{\mathbf{x}}  \\ \ddot{\mathbf{x}}
    \end{bmatrix}=\begin{bmatrix}
     \mathbf{0} & \mathbf{I} \\ \mathbf{0}  & \mathbf{0}
    \end{bmatrix}\begin{bmatrix}
      \mathbf{x}  \\\dot{\mathbf{x}}
    \end{bmatrix}+\begin{bmatrix}
       \mathbf{0} \\ (-\mathbf{D}^\top\otimes \mathbf{I}_d)
    \end{bmatrix}\mathbf{u}.
\end{align}
We can rewrite \(\mathbf{u}\) in terms of \(\mathbf{x}\) and \(\dot{\mathbf{x}}\) as
\begin{align}\label{eq:control_edge}
    \mathbf{u}&=\begin{bmatrix}
      \mathbf{D}^s  & \mathbf{D}^a
    \end{bmatrix}\begin{bmatrix}
      \mathbf{x}  \\ \dot{\mathbf{x}}
    \end{bmatrix}.
\end{align}
By substituting \eqref{eq:control_edge} into the edge interaction dynamics \eqref{eq:edge-dynamics}, the resulting closed-loop system is then of the form
\begin{align*}
    \begin{bmatrix}
      \dot{\mathbf{x}}  \\ \ddot{\mathbf{x}}
    \end{bmatrix}&=\begin{bmatrix}
     \mathbf{0} & \mathbf{I} \\ -(\mathbf{D}^\top\otimes \mathbf{I}_d)\mathbf{D}^s & -(\mathbf{D}^\top\otimes \mathbf{I}_d)\mathbf{D}^a
    \end{bmatrix}\begin{bmatrix}
      \mathbf{x}  \\\dot{\mathbf{x}}
    \end{bmatrix},
\end{align*}
where \((\mathbf{D}^\top\otimes \mathbf{I}_d)\mathbf{D}^s\) and \((\mathbf{D}^\top\otimes \mathbf{I}_d)\mathbf{D}^a\) are the edge Laplacian-alike matrices of the directed graph \(\mathcal{G}\) and positive definite when \(\mathcal{G}\) is spanning tree~\cite{graph_methods_book}.  

The eigenvalues of the closed-loop system matrix 
\[
\mathbf{A}_{cl}=\begin{bmatrix}
     \mathbf{0} & \mathbf{I} \\ -(\mathbf{D}^\top\otimes \mathbf{I}_d)\mathbf{D}^s & -(\mathbf{D}^\top\otimes \mathbf{I}_d)\mathbf{D}^a
    \end{bmatrix},
\]
satisfy the characteristic equation
\begin{equation}\label{eq:charact}
\text{det}(\lambda\mathbf{I}-\mathbf{A}_{cl})=\text{det}(\lambda^2\mathbf{I}+\lambda(\mathbf{D}^\top\otimes \mathbf{I}_d)\mathbf{D}^a+(\mathbf{D}^\top\otimes \mathbf{I}_d)\mathbf{D}^s)=0.
\end{equation}

Denoting the eigenvalues of \((\mathbf{D}^\top\otimes \mathbf{I}_d)\mathbf{D}^a\) and \((\mathbf{D}^\top\otimes \mathbf{I}_d)\mathbf{D}^s\) as \(\mu_i\) and \(\gamma_i\), respectively, the characteristic equation~\eqref{eq:charact} implies that for every \(i\) the eigenvalue \( \lambda_i \) satisfies 
\[
\lambda_i^2 + \lambda_i \mu_i + \gamma_i = 0,
\]
or
\[
\lambda_i = \frac{1}{2}(-\mu_i \pm \sqrt{\mu_i^2 - 4\gamma_i}).
\]

The Laplacian-alike matrices \((\mathbf{D}^\top\otimes \mathbf{I}_d)\mathbf{D}^a\) and \((\mathbf{D}^\top\otimes \mathbf{I}_d)\mathbf{D}^s\) inherit the eigenvalue property of the Laplacian matrix, i.e., their all eigenvalues have nonnegative real parts. As a result, the eigenvalues \( \lambda_i \) of the closed-loop system matrix will have negative real parts. Consequently, the matrix \(\mathbf{A}_{cl}\) is Hurwitz. This guarantees~\eqref{eq:AS}.
\end{proof}

From Theorem~\ref{theorem:stability}, it follows that as long as the interaction graph \(\mathcal{G}\) forms a spanning tree, the MAS with the designed local control strategies will be asymptotically stable for any initial condition. This condition ensures that, regardless of the initial states of the agents, the MAS will eventually converge to a stable configuration in which the agents exhibit the desired flocking behavior.

\section{Simulation Results}\label{sec:sim}
This section presents illustrative examples to validate the bearing-distance flocking model in Section~\ref {sec:bearing}. Before, we introduce three metrics to analyze the evolution and maintenance of coordination, cohesion, and collision avoidance. We quantify coordination using a metric that evaluates the overall alignment by computing the average cosine of the angles between all pairs of unit velocity vectors. The coordination metric \(C\) is defined as,  
\[
C = \text{avg}(\sum_{\forall(i,j)} \frac{\mathbf{v}_i^\top \mathbf{v}_j}{\|\mathbf{v}_i\| \| \mathbf{v}_j\|}),
\]
where \(\text{avg}(\cdot)\) denotes the average value, and the dot product returns the cosine of the angle between the unit velocity vectors. The maximum value of \(C = 1\) signifies perfect alignment, representing ideal coordination. As \(C\) decreases from its maximum value, it indicates a progressive loss of coordination, whereas an increase toward the maximum value reflects a progressive improvement in coordination. To demonstrate the cohesive movement of the flocking group, a cohesive radius is defined as the maximum distance between the flocking center and an agent as~\cite{4895844},
\[
R = \max_{i} \| \mathbf{p}_i - \frac{1}{n} \sum_{i=1}^{n} \mathbf{p}_i \|.
\]  
If the minimum distance between any two agents, defined as  
\[
D = \min_{\forall(i,j)} \| \mathbf{p}_j - \mathbf{p}_i \|,
\]  
remains greater than a predefined safe separation threshold during flocking, it ensures that no collisions occur among agents.

In an unconstrained environment, free of dynamic and static obstacles or boundary restrictions, the metrics of coordination, cohesive radius, and minimum distance must remain stable during flocking. Stability in these metrics signifies that the flocking group maintains perfect order and connectivity, avoiding the separation of agents from the group or fragmentation of the flock while ensuring that no inter-agent collisions occur. The presence of environmental hazards can disrupt the stability of the flock and cause the separation of agents or fragmentation of the flock. 

In the following subsections, we present illustrative examples of flocking in an unconstrained environment and a constrained environment. The constrained environment includes an alien agent, an irregularly shaped stationary obstacle, and is confined. Simulation examples include a MAS with \(n = 10\) agents, indexed by \(\mathcal{V} = \{1, \cdots, 10\}\), operating in a 2D environment (\(d = 2\)). 

\subsection{Unconstrained Environment}
 
For each agent \(i \in \mathcal{V}\), the zone radii are set as follows: the repulsion zone radius \(r_i = 1.5\) \(\mathrm{m}\), the conflict zone radius \(c_i = 3.0\) \(\mathrm{m}\), and the attraction zone radius \(a_i = 5.0\) \(\mathrm{m}\). We set weights as follows: repulsion weights \(\omega_{ij}^{ls} = 5\), cohesion attraction weights \(\omega_{ij}^{lc} = 0.75\), and alignment attraction weights \(\omega_{ij}^{la} = 0.25\). The global attraction weights for each agent are defined as \(\omega_i^{ga} = 5\). Repulsion weights are larger than the attraction weight to prioritize separation and collision avoidance over cohesion and alignment. The initial positions and velocities of the agents are random within the range $(0,10)$ \(\mathrm{m}\) and \((0,1)\) \(\mathrm{m/s}\). The maximum speed and acceleration for each agent are set to \(v_i^{\text{max}} = 4\) \(\mathrm{m/s}\) and \(u_i^{\text{max}} = 2\) \(\mathrm{m/s^2}\), respectively. The desired global flock speed is defined as \(v_d = 3\) \(\mathrm{m/s}\). The total time is $20.0$ \(\mathrm{s}\), with a time step of $0.1$ \(\mathrm{s}\).

\begin{figure*}[t]
\centering
    % First row
    \begin{subfigure}{0.24\linewidth}
        \centering
        \includegraphics[width=\linewidth]{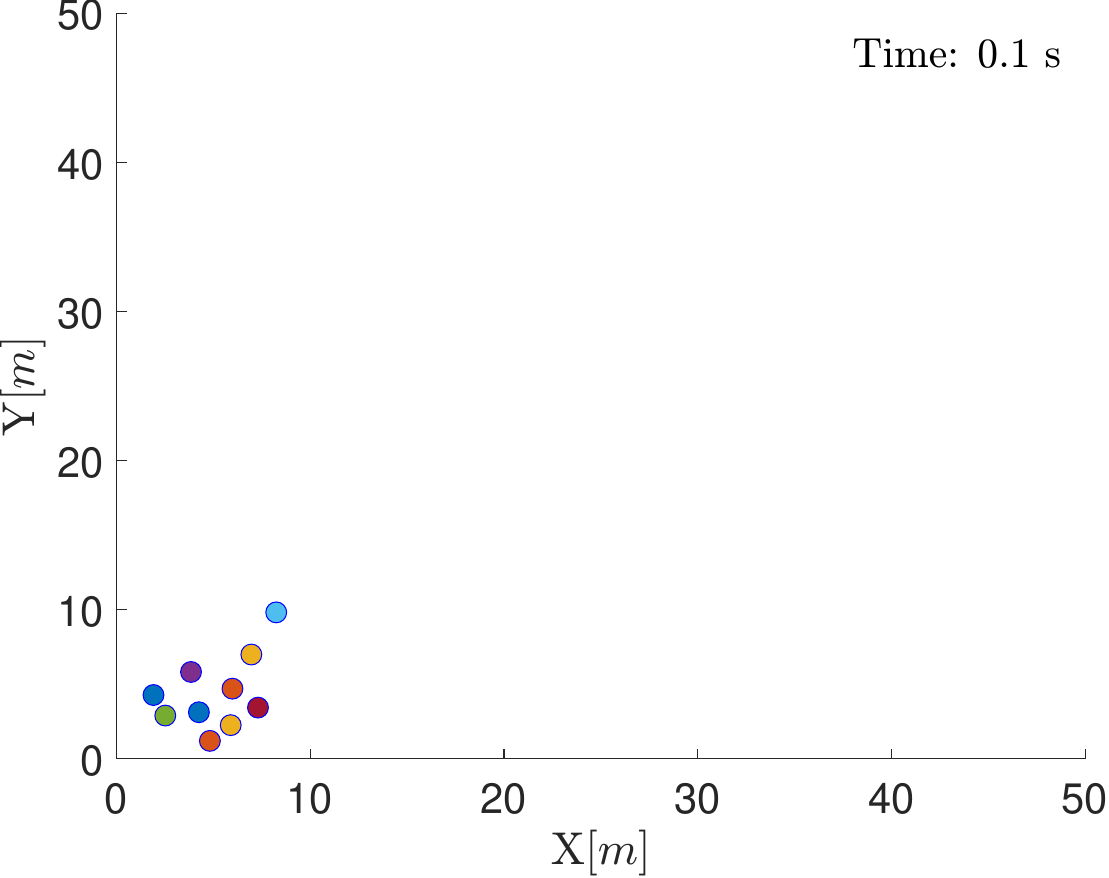}
        \subcaption[]{\(t=0.1\) s} \label{fig:C}
    \end{subfigure}
    \hfill
    \begin{subfigure}{0.24\linewidth}
        \centering
        \includegraphics[width=\linewidth]{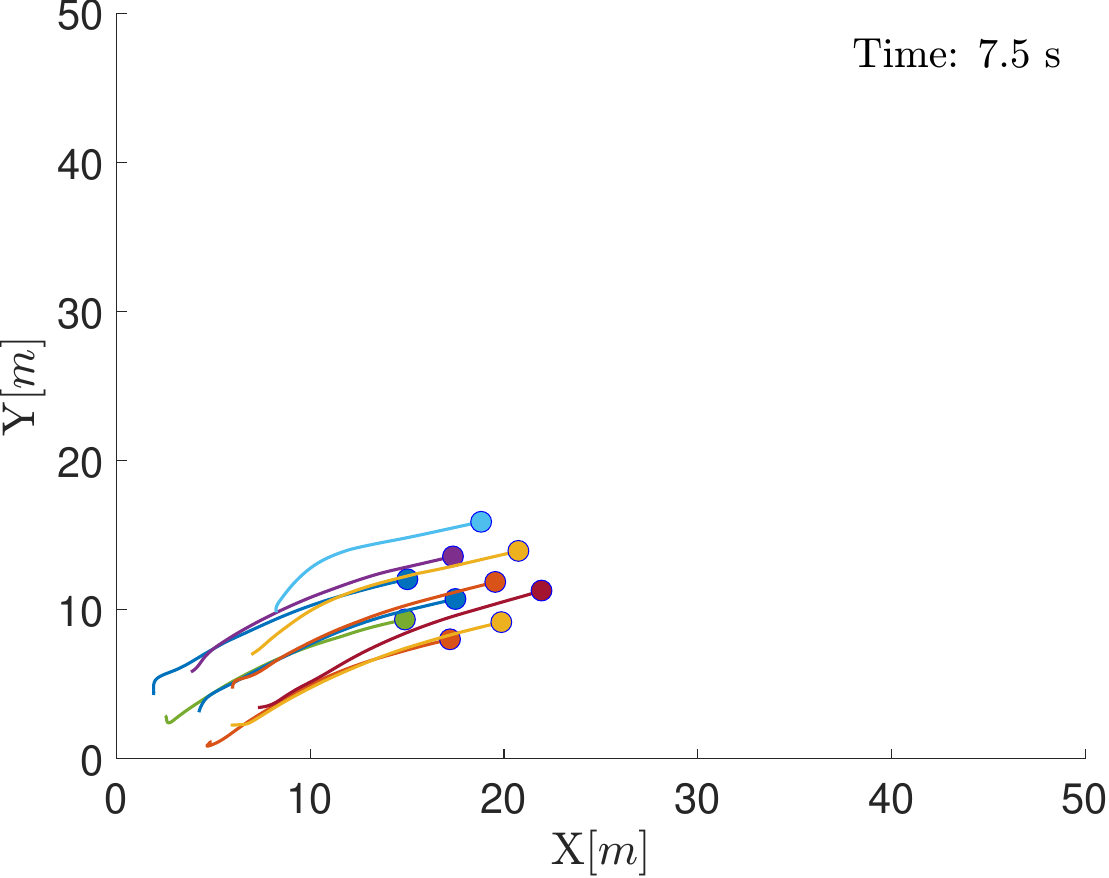}
        \subcaption[]{\(t=7.5\) s}
    \end{subfigure}
    \hfill
    \begin{subfigure}{0.24\linewidth}
        \centering
        \includegraphics[width=\linewidth]{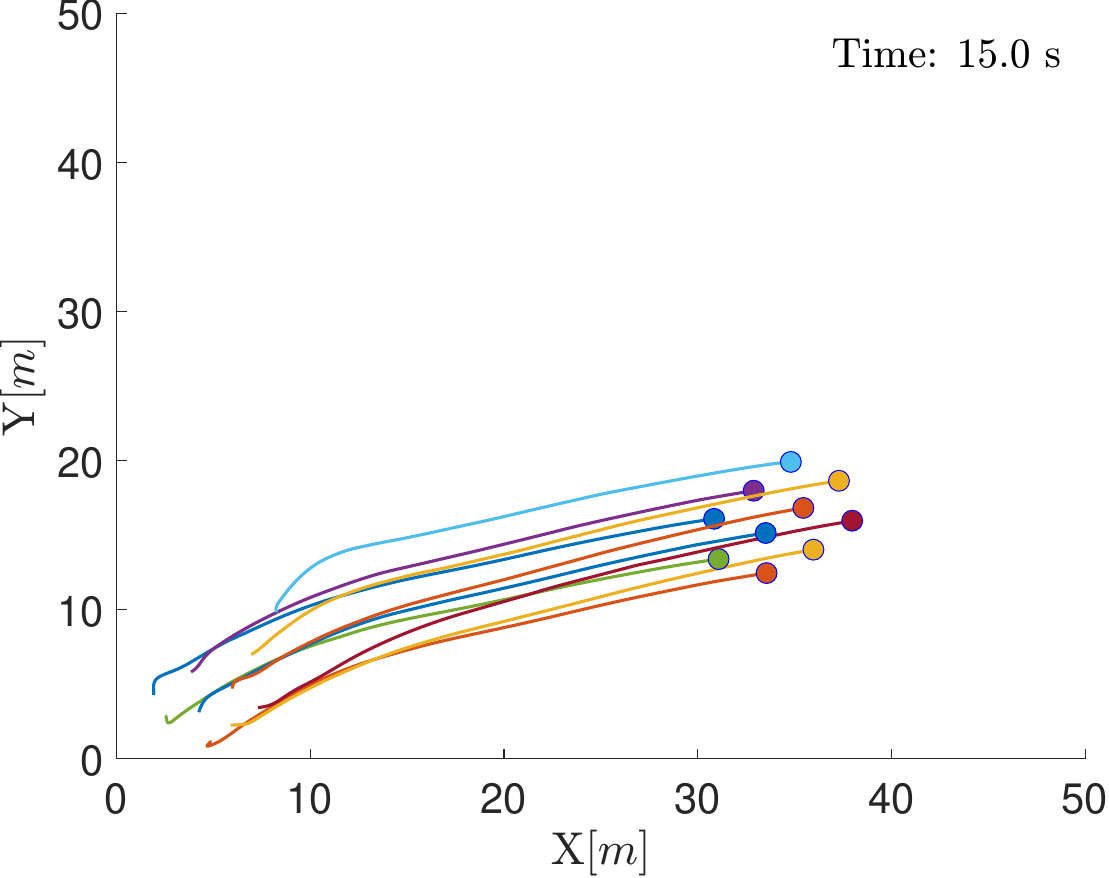}
        \subcaption[]{\(t=15.0\) s} 
    \end{subfigure}
    \hfill
    \begin{subfigure}{0.24\linewidth}
        \centering
        \includegraphics[width=\linewidth]{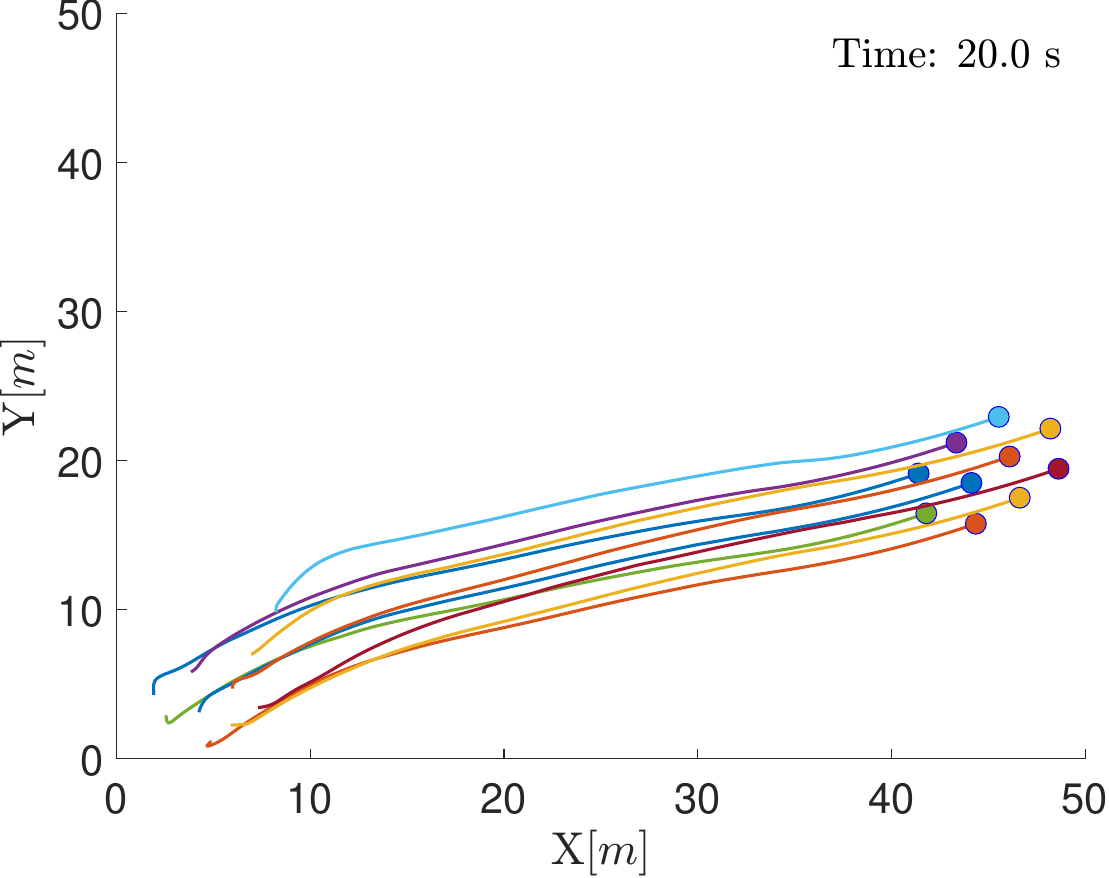}
        \subcaption[]{\(t=20.0\) s} 
    \end{subfigure}

   \caption{The flocking behavior of \(10\) agents in an unconstrained environment exhibits cohesive, collision-free collective motion.}
   \label{fig:uncon}
\end{figure*}

\begin{figure*}[t]
   \centering
    % First row
    \begin{subfigure}{0.32\linewidth}
        \centering
        \includegraphics[width=\linewidth]{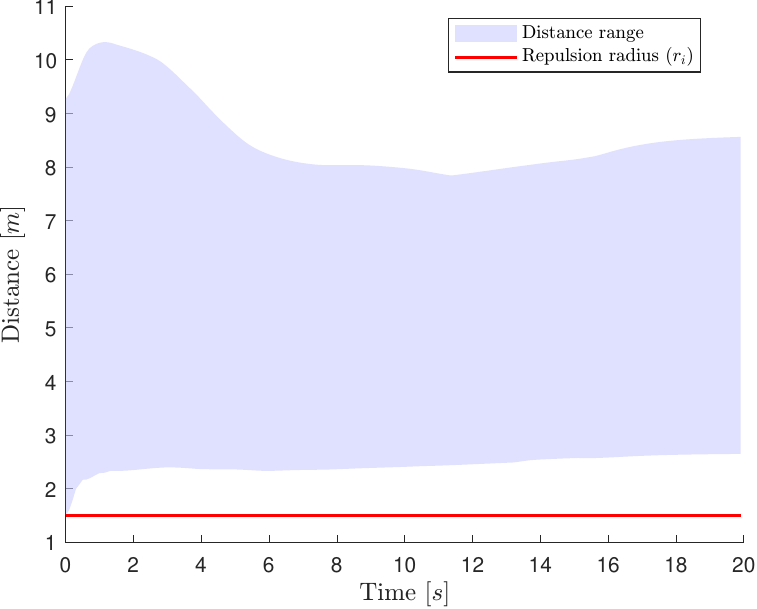}
        \subcaption[]{}
    \end{subfigure}
    \hfill
    \begin{subfigure}{0.32\linewidth}
        \centering
        \includegraphics[width=\linewidth]{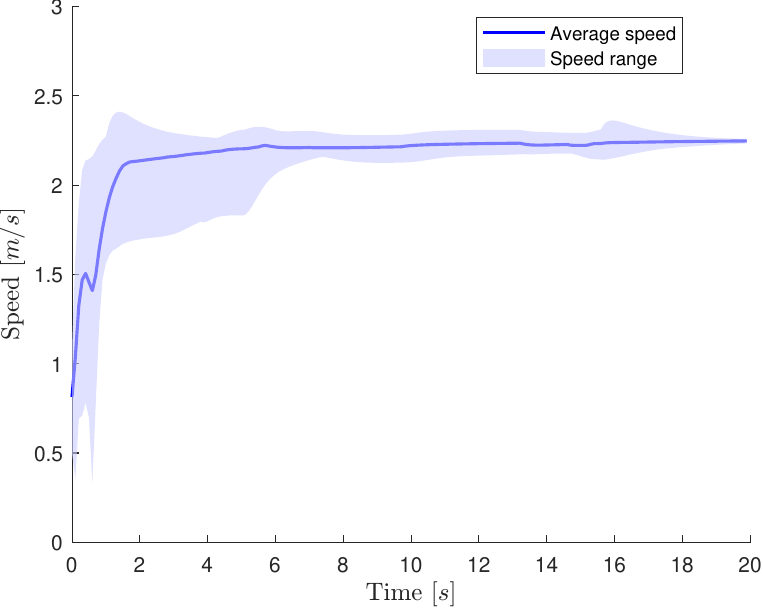}
        \subcaption[]{} 
    \end{subfigure}
    \hfill
    \begin{subfigure}{0.32\linewidth}
        \centering
        \includegraphics[width=\linewidth]{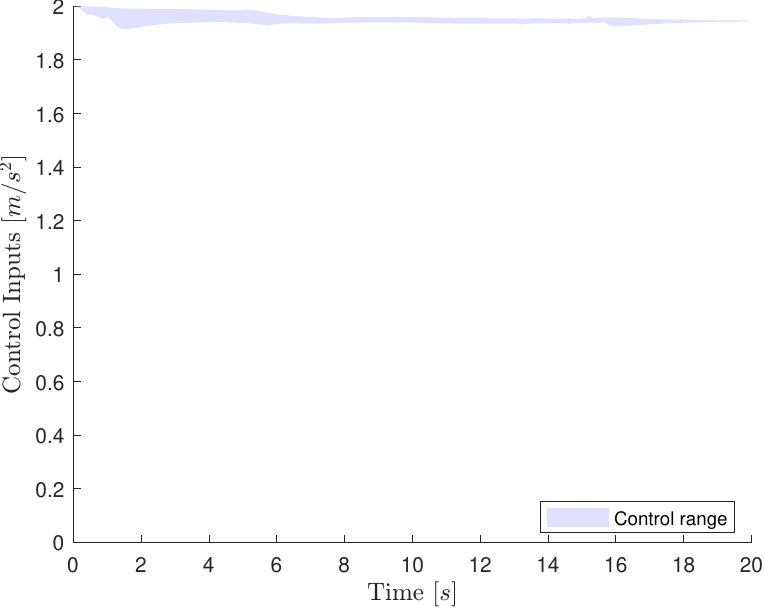}
        \subcaption[]{} 
    \end{subfigure}
\caption{Inter-agent distances, speeds, and control input profiles of the flock agents. The inter-agent distances remain above the individual repulsion radii ($r_i=1.5$ \(\mathrm{m}\) for all $i$). Agent speeds are below the maximum speeds (\( v_i^{\text{max}} = 4\) \(\mathrm{m/s}\) for all \( i \)), and the average flock speed stays close to the desired global speed (\( v_d = 3\) \(\mathrm{m/s}\) ). Control inputs are within their bounds (\( u_i^{\text{max}} = 2\) \(\mathrm{m/s^2}\) for all \( i \)).}
\label{fig:profile_uncon}
\end{figure*}

\begin{figure}[ht]
   \centering
    % First row
    \begin{subfigure}{0.32\linewidth}
        \centering
        \includegraphics[width=\linewidth]{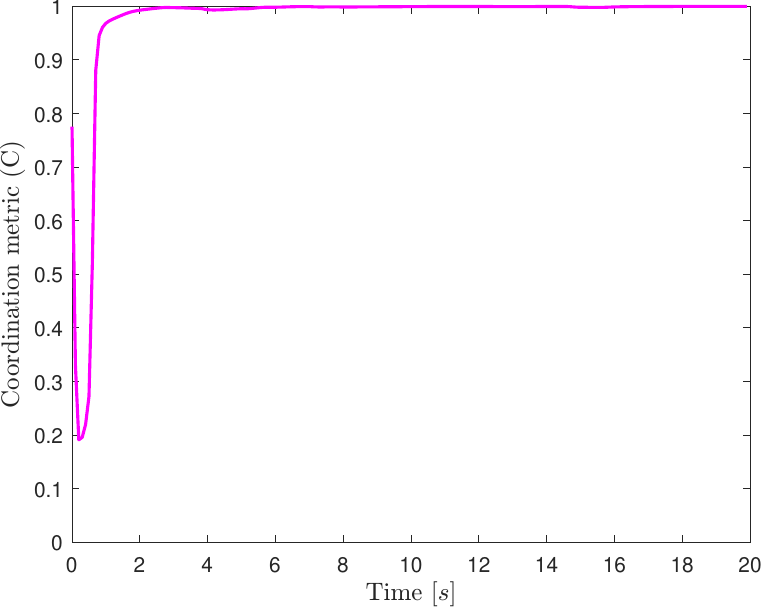}
        \subcaption[]{} \label{fig:C_uncon}
    \end{subfigure}
    \hfill
    \begin{subfigure}{0.32\linewidth}
        \centering
        \includegraphics[width=\linewidth]{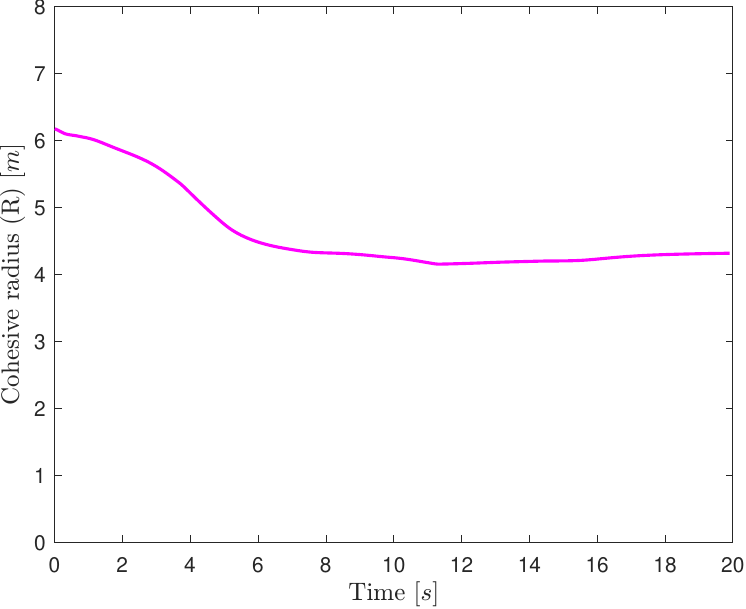}
        \subcaption[]{} \label{fig:R_uncon}
    \end{subfigure}
    \hfill
    \begin{subfigure}{0.32\linewidth}
        \centering
        \includegraphics[width=\linewidth]{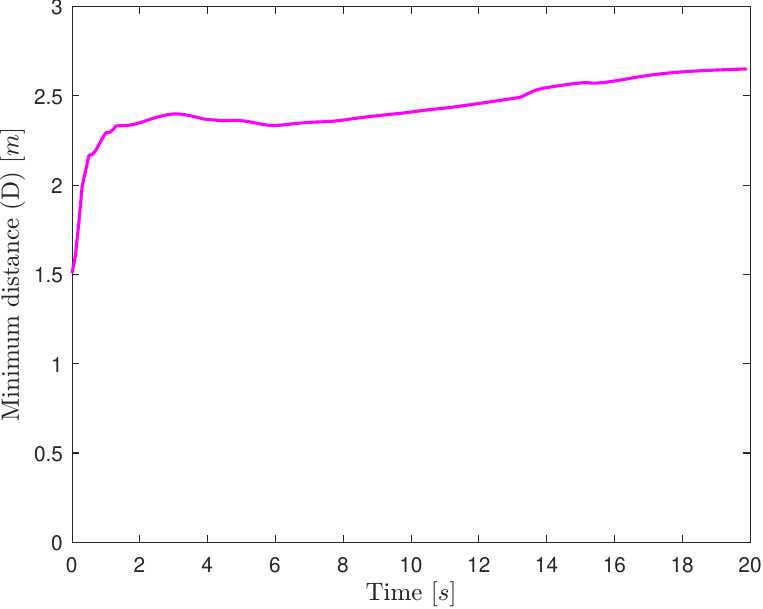}
        \subcaption[]{} \label{fig:D_uncon}
    \end{subfigure}
\caption{Time histories of (a) coordination metric, (b) cohesive radius, and (c) minimum distance.}
\label{fig:metrics-uncon}
\end{figure}

Fig.~\ref{fig:uncon} illustrates the resultant flocking behavior. It is observed that in an unconstrained environment, the agents successfully demonstrate cohesive motion, maintaining a stable spatial structure as a unified group. Fig.~\ref{fig:profile_uncon} shows the profiles of inter-agent distances, speeds, and control inputs. The distances between agents stay strictly above the repulsion radius of \(r_i = 1.5\) \(\mathrm{m}\). Agent speeds remain below the maximum of \(4\) \(\mathrm{m/s}\), with the average flock speed close to the desired \(3\) \(\mathrm{m/s}\). The control inputs are within the specified limits of \(2\) \(\mathrm{m/s^2}\). Furthermore, the time histories of coordination, cohesion, and collision avoidance metrics are depicted in Fig.~\ref{fig:metrics-uncon}. As shown in Fig.~\ref{fig:metrics-uncon}(\subref{fig:C_uncon}), a perfect motion direction order is achieved within \(2\) \(\mathrm{s}\) and is maintained for the rest of the time. The time history of the cohesive radius in Fig.~\ref{fig:metrics-uncon}(\subref{fig:R_uncon}) demonstrates that it remained stable once the flock itself had stabilized around \(t = 5\) \(\mathrm{s}\). Finally, as shown in Fig.~\ref{fig:metrics-uncon}(\subref{fig:D_uncon}), the minimum inter-agent distance remains above the repulsion radius (\(r_i = 1.5\) \(\mathrm{m}\)), ensuring collision-free interactions throughout the process.

\subsection{Constrained Dynamic Environment}

We consider a constrained dynamic environment that includes an alien agent, a stationary obstacle, and boundaries. The initialization follows the same setup as in the previous example unless otherwise stated here. For each agent \(i \in \mathcal{V}\), the surveillance zone radius is \(s_i = 6.0\) \(\mathrm{m}\). The surveillance zone repulsion weight \(\omega_{ik}^{ss}\) is set to 5. An alien agent is trapped within a triangular zone within the environment. The flock agents are unaware of the alien's presence until it enters their surveillance zone. The alien can detect flock agents within a radius of $9$ \(\mathrm{m}\). Once the flock agents come within range, the alien follows the closest agent until it reaches the borders of its triangular zone. The shorter surveillance radius of the flock agents causes a delayed reaction to the presence of the alien. The environment also contains an irregularly shaped stationary obstacle. The obstacle avoidance weight \(\omega_i^{oa}\) is set to 5. The environment is confined, with boundaries from $[0,0]$ \(\mathrm{m}\) to $[50,50]$ \(\mathrm{m}\). The total time is $100.0$ \(\mathrm{s}\).

\begin{figure*}[h]
\centering
    % First row
    \begin{subfigure}{0.24\linewidth}
        \centering
        \includegraphics[width=\linewidth]{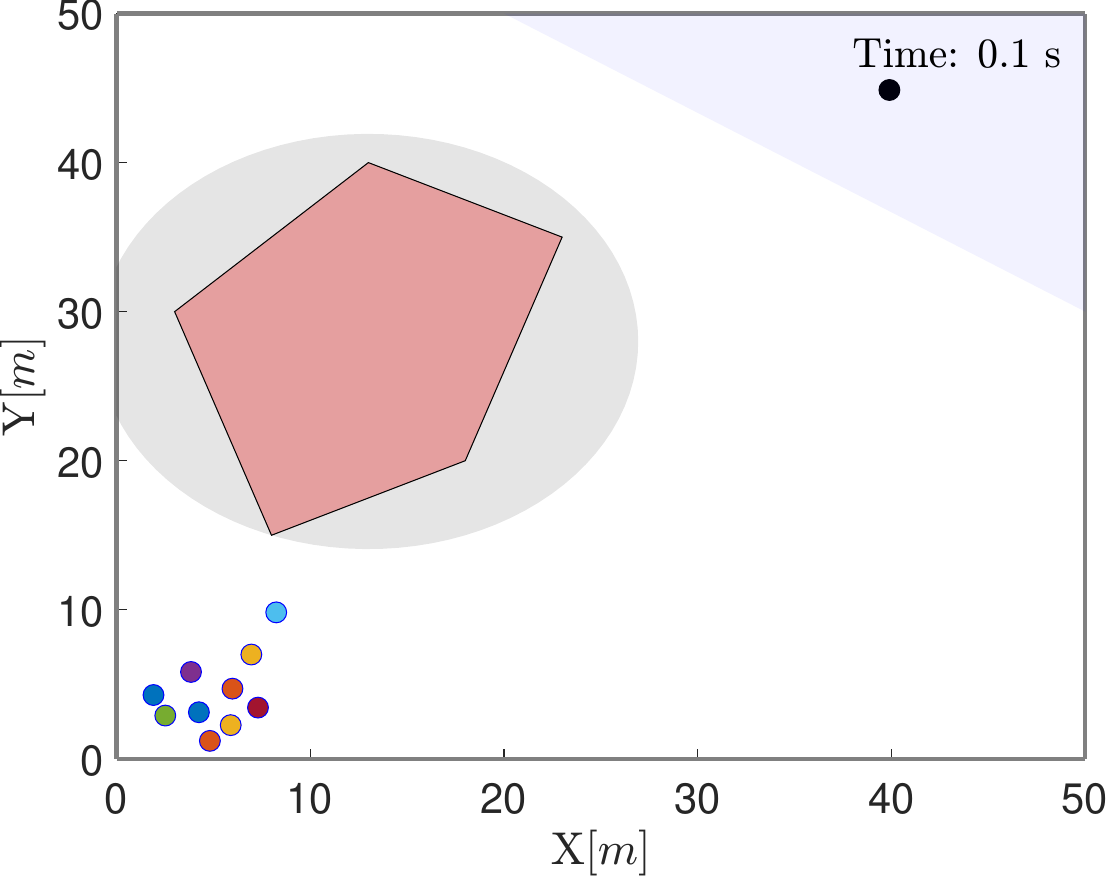}
        \subcaption[]{\(t=0.1\) s} 
    \end{subfigure}
    \hfill
    \begin{subfigure}{0.24\linewidth}
        \centering
        \includegraphics[width=\linewidth]{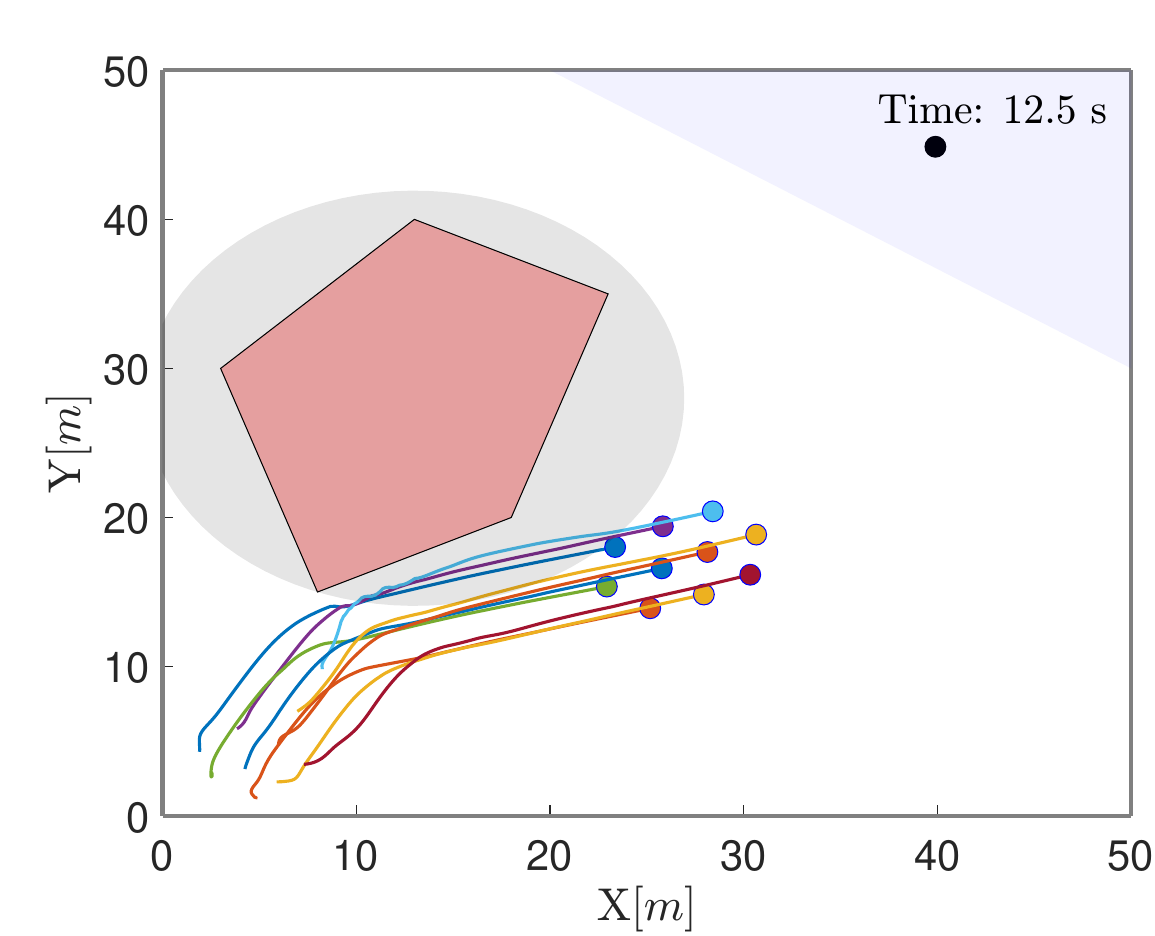}
        \subcaption[]{\(t=12.5\) s} 
    \end{subfigure}
    \hfill
    \begin{subfigure}{0.24\linewidth}
        \centering
        \includegraphics[width=\linewidth]{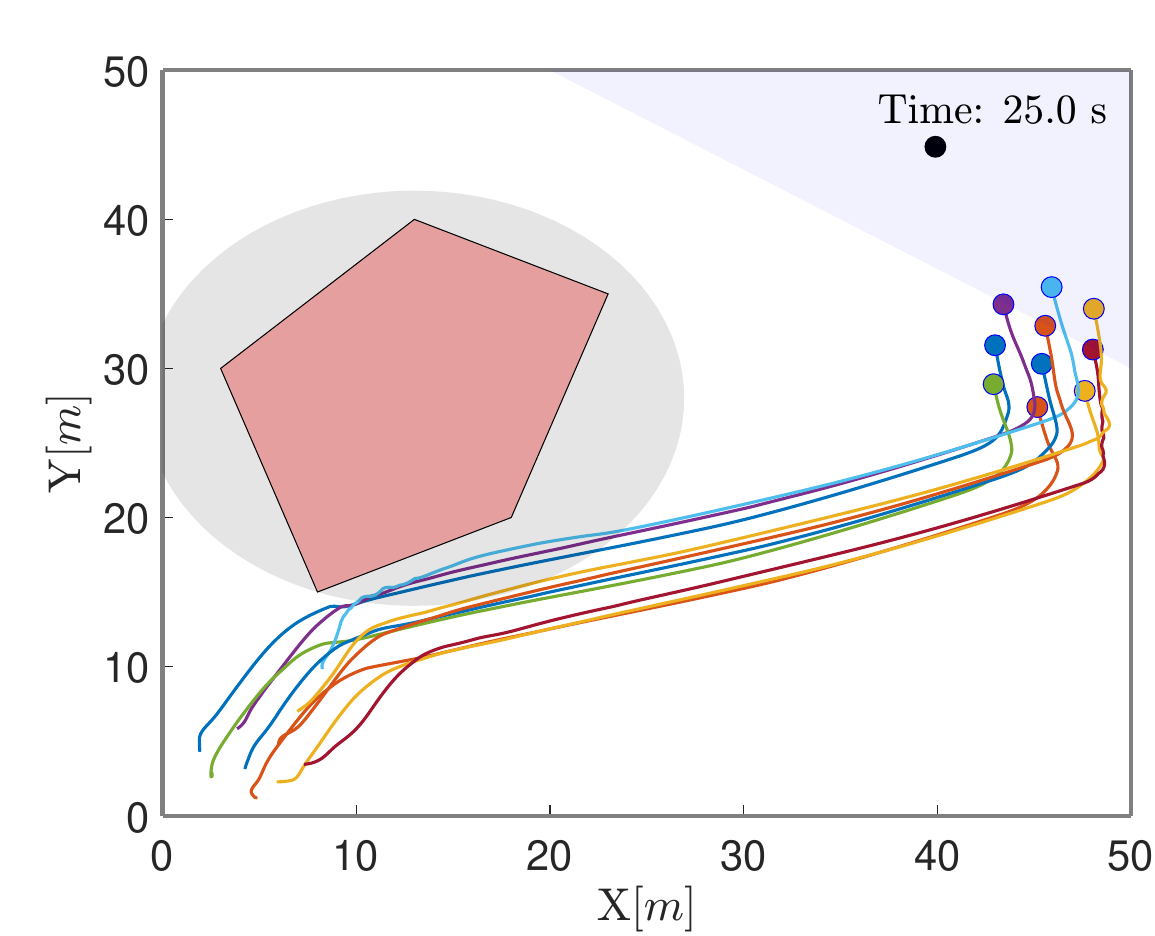}
        \subcaption[]{\(t=25.0\) s} 
    \end{subfigure}
    \hfill
    \begin{subfigure}{0.24\linewidth}
        \centering
        \includegraphics[width=\linewidth]{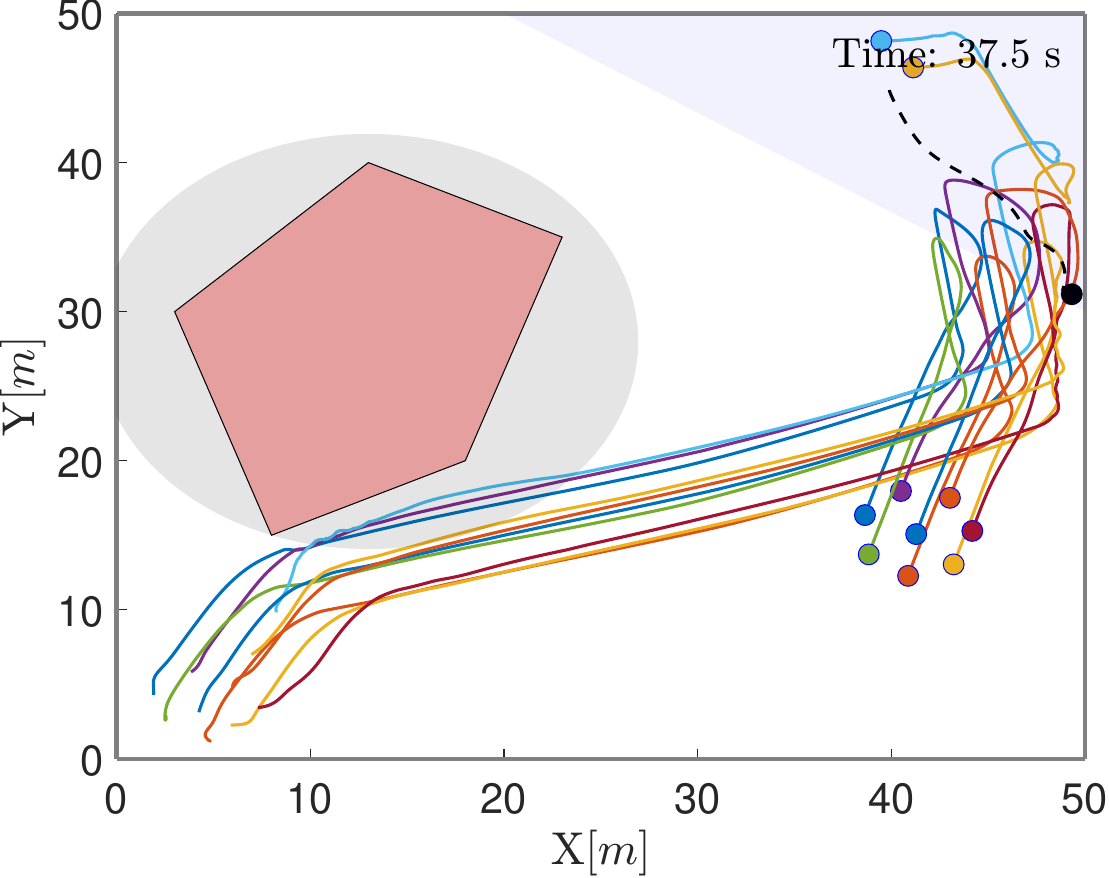}
        \subcaption[]{\(t=37.5\) s}
    \end{subfigure}
    % Second row
    \vskip 0.5\baselineskip
    \begin{subfigure}{0.24\linewidth}
        \centering
        \includegraphics[width=\linewidth]{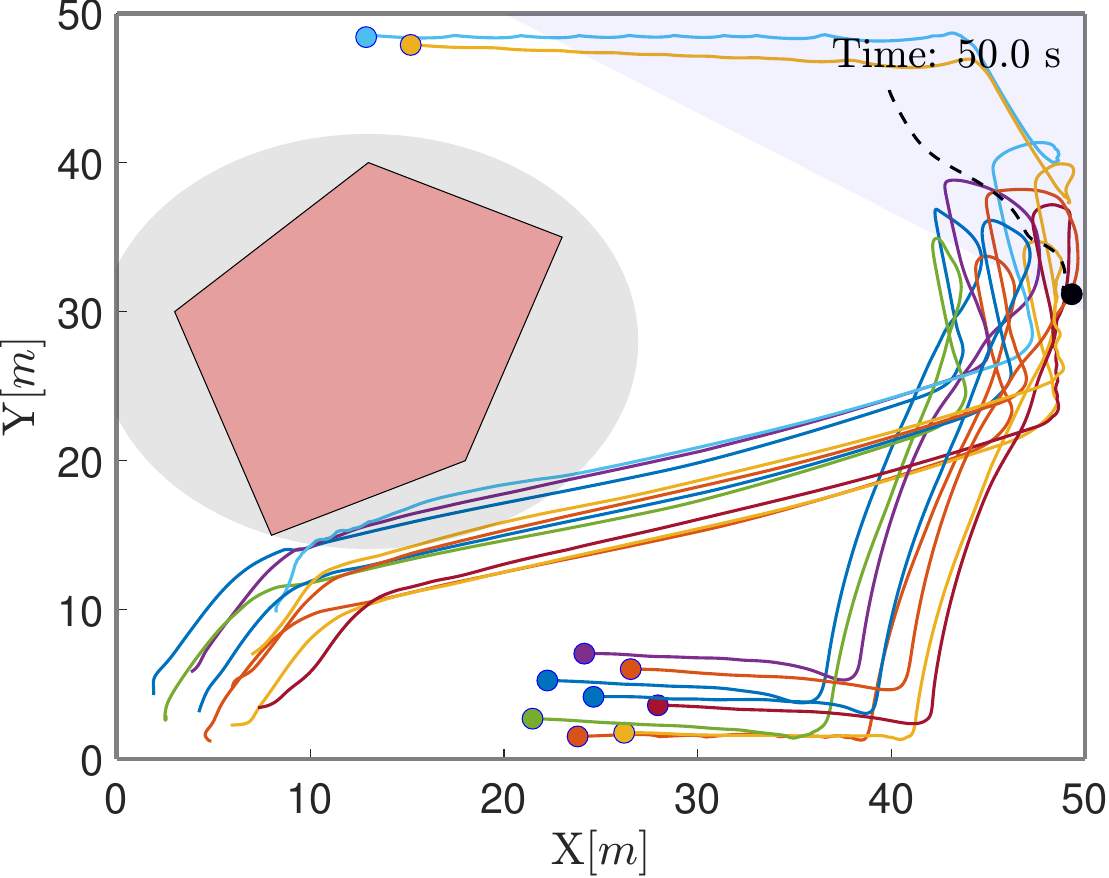}
        \subcaption[]{\(t=50.0\) s} 
    \end{subfigure}
    \hfill
    \begin{subfigure}{0.24\linewidth}
        \centering
        \includegraphics[width=\linewidth]{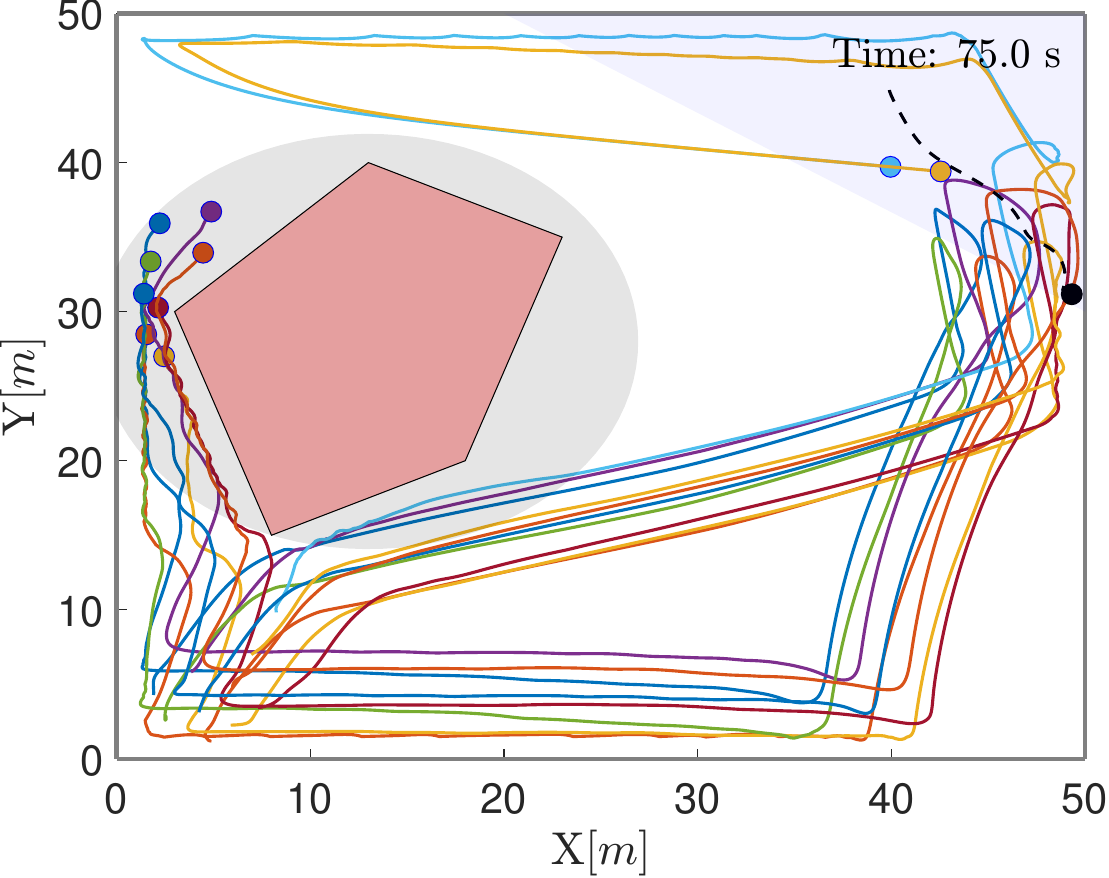}
        \subcaption[]{\(t=75.0\) s}
    \end{subfigure}
    \hfill
    \begin{subfigure}{0.24\linewidth}
        \centering
        \includegraphics[width=\linewidth]{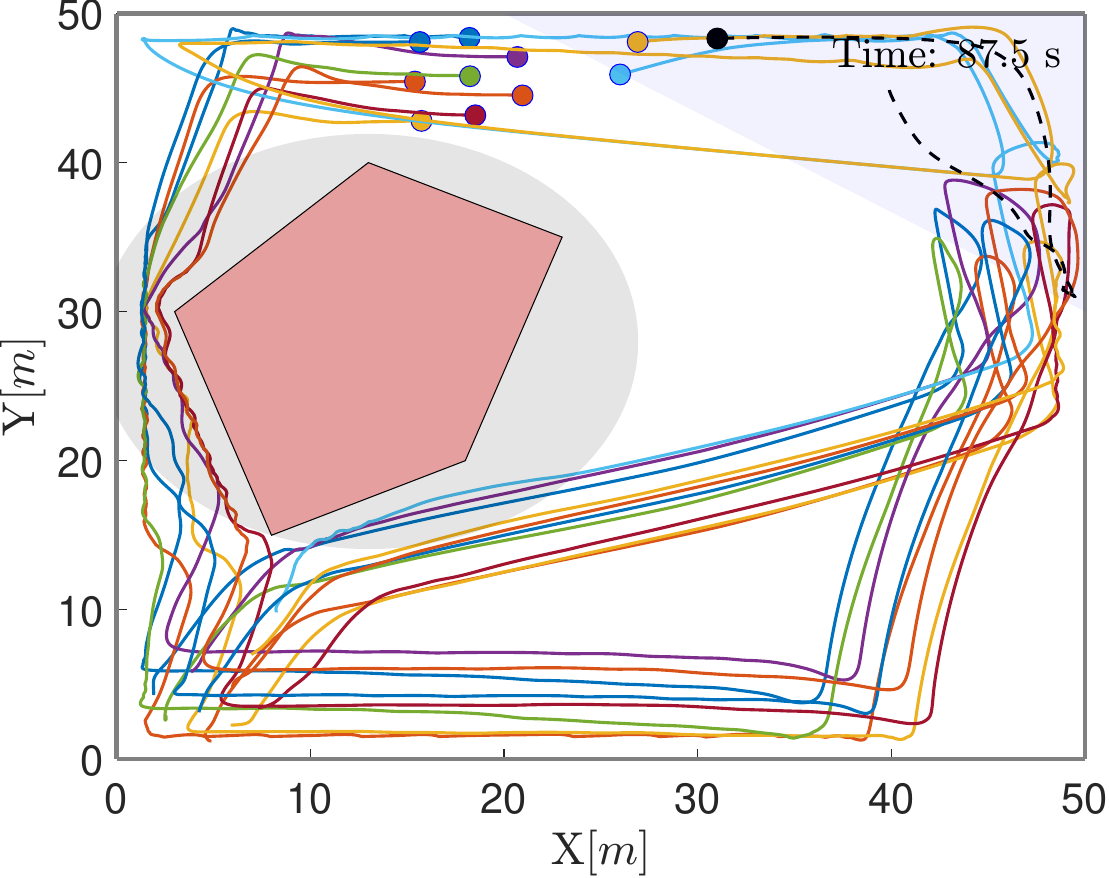}
        \subcaption[]{\(t=87.5\) s}
    \end{subfigure}
    \hfill
    \begin{subfigure}{0.24\linewidth}
        \centering
        \includegraphics[width=\linewidth]{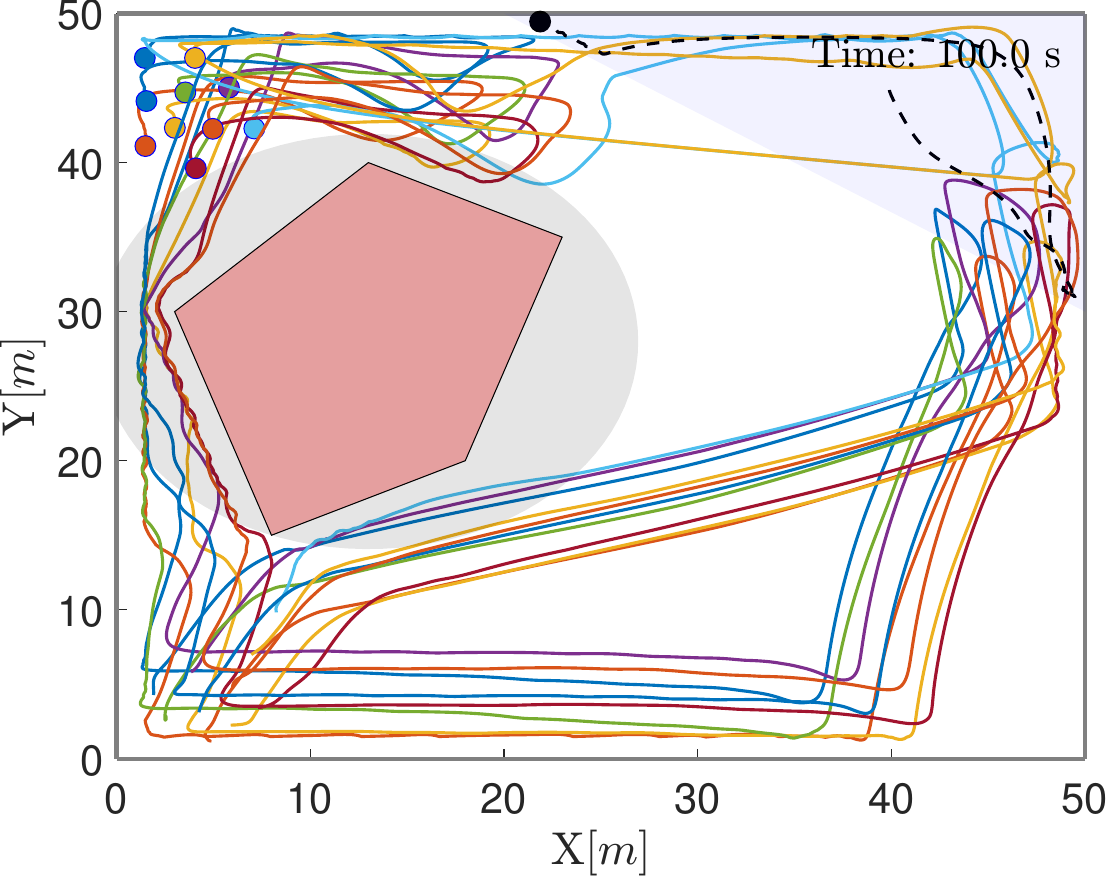}
        \subcaption[]{\(t=100.0\) s} 
    \end{subfigure}
   \caption{Flocking behavior of \( 10 \) agents in a dynamic environment. The flocking environment is confined and includes an irregular polygon-shaped stationary obstacle and an alien agent. The alien (shown in black), constrained within the transparent blue zone, begins to follow the nearest flocking agent when within proximity. The flock initially demonstrates cohesive and collision-free motion; however, the arrival of the alien agent causes a temporary fragmentation, though the flock eventually reunites. The agents navigate through a narrow passage between the stationary obstacle and the environmental boundaries. Such a maneuver would not be permitted with a conventional obstacle avoidance method, which employs a circular buffer zone around irregular obstacles (depicted in gray).}
   \label{fig:con}
\end{figure*}

\begin{figure*}[htbp]
   \centering
    % First row
    \begin{subfigure}{0.32\linewidth}
        \centering
        \includegraphics[width=\linewidth]{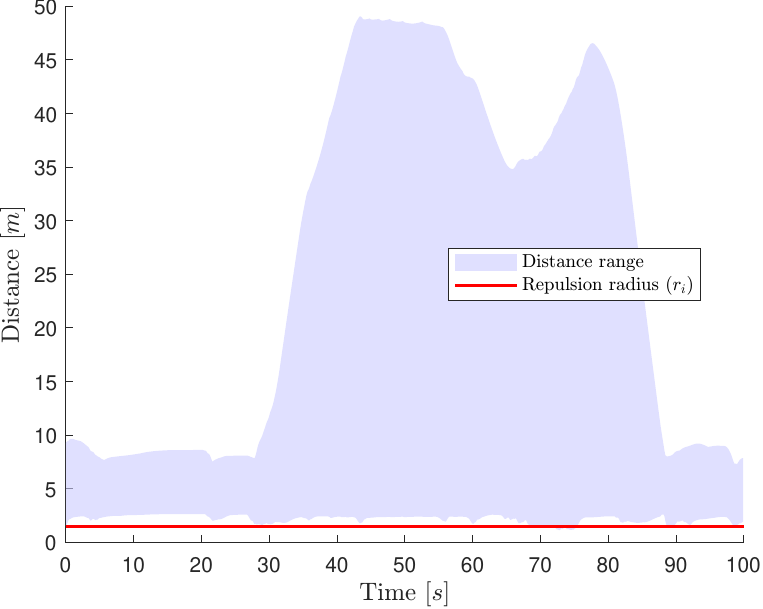}
        \subcaption[]{} \label{fig:dist_con}
    \end{subfigure}
    \hfill
    \begin{subfigure}{0.32\linewidth}
        \centering
        \includegraphics[width=\linewidth]{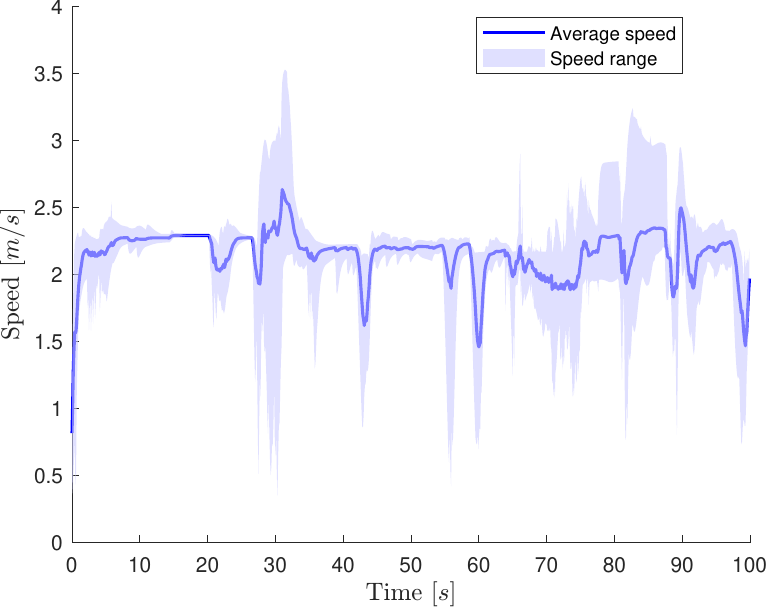}
        \subcaption[]{} \label{fig:speed_con}
    \end{subfigure}
    \hfill
    \begin{subfigure}{0.32\linewidth}
        \centering
        \includegraphics[width=\linewidth]{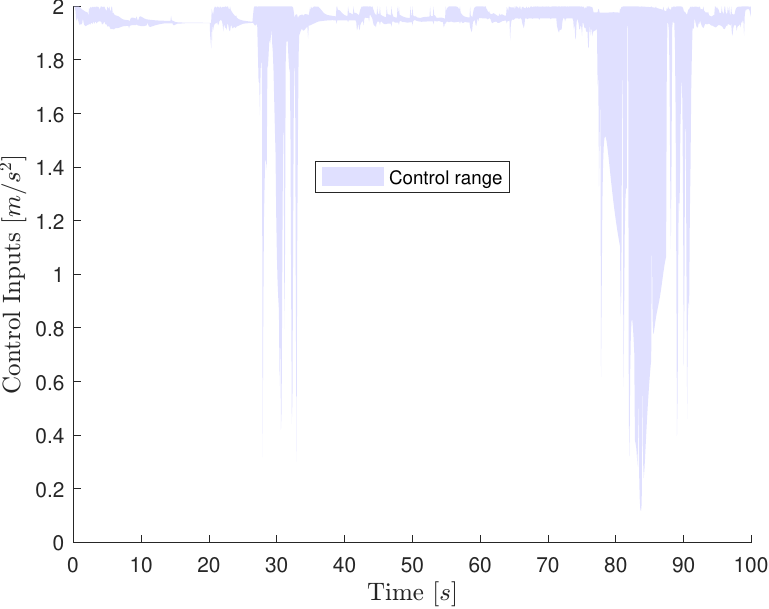}
        \subcaption[]{} \label{fig:cont_con}
    \end{subfigure}
\caption{Inter-agent distances, speeds, and control input profiles of the flocking agents in the constrained dynamic environment. The inter-agent distances consistently remain above the individual repulsion radii. When the flock is cohesive, the distances range up to 10 m. Fragmentation occurs due to the alien's interference, causing a significant increase in distance range before returning to the initial range upon reunification. Agent speeds are below the maximum, and the average flock speed remains near the desired global speed, with control inputs within their bounds.}
\label{fig:profile_con}
\end{figure*}

Fig.~\ref{fig:con} illustrates the behavior of the agents. The alien agent follows the nearest flock agent when close, causing temporary fragmentation in the cohesive flock. Agents navigate a narrow passage between the obstacle and boundaries, a maneuver infeasible with conventional circular buffer zone methods. Fig.~\ref{fig:profile_con} shows the profiles of inter-agent distances, speeds, and control inputs. Inter-agent distances stay above the repulsion radius of \(r_i = 1.5\) \(\mathrm{m}\), reaching up to 10 meters when cohesive. Fragmentation due to the alien’s interference temporarily increases distance ranges before normalizing. Agent speeds remain below the maximum of \(4\) \(\mathrm{m/s}\), with the average flock speed close to the desired \(3\) \(\mathrm{m/s}\). Control inputs are within the specified bounds of \(2\) \(\mathrm{m/s^2}\). Furthermore, from the behaviors of the agents depicted in Fig.~\ref{fig:con}, it is clear that flock coordination and cohesion have not been maintained due to fragmentation of the flock. However, collision avoidance is still preserved, as evident from Fig.~\ref{fig:con}(\subref{fig:dist_con}).

The simulation examples illustrated the emergence of flocking behavior in accordance with the zone-based flocking rules defined in subsection~\ref{sec:rules}. The agents, initially positioned and moving randomly, formed a cohesive group, maintaining relative distances while moving as a collective entity. The flocking group, when placed within a constrained environment, effectively navigated around an irregular obstacle with the proposed directionally aware obstacle avoidance mechanism, which allowed closer proximity to the obstacle's edges and passage through a narrow gap, maneuvers not feasible with conventional circular buffer zone methods. Despite temporary fragmentation caused by an alien agent, the flock reformed cohesively. These results highlight the effectiveness of the bearing-distance flocking approach with zone-based interactions for realistic \textit{boids}-like flocking.

\begin{figure*}[ht]
      \includegraphics[width=0.32\textwidth]{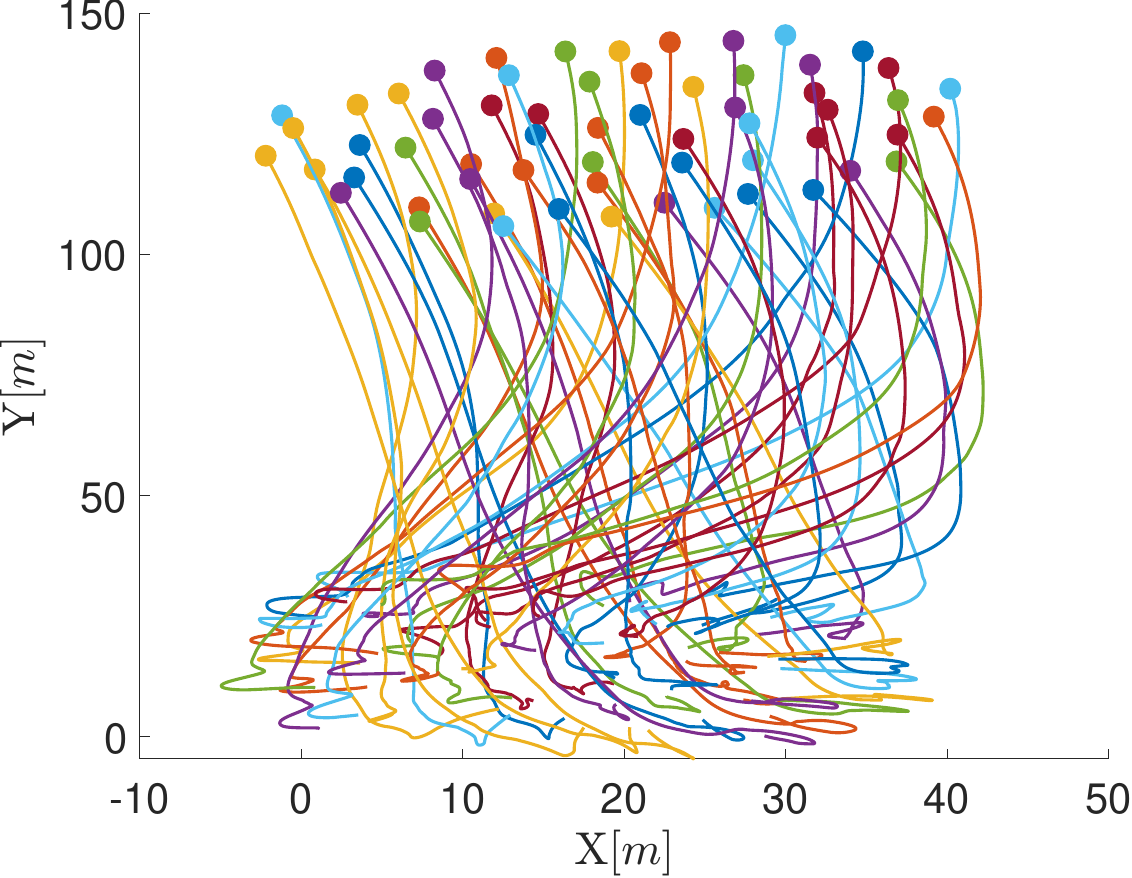}
      \includegraphics[width=0.32\textwidth]{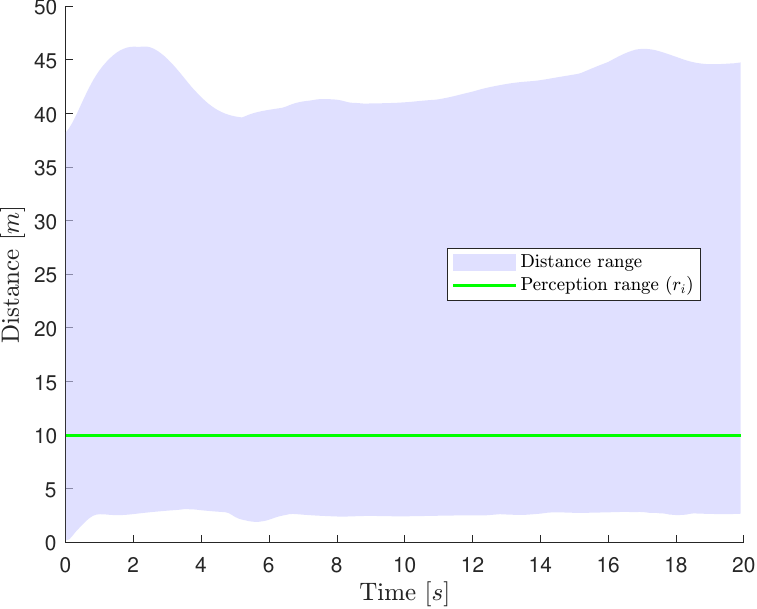}
      \includegraphics[width=0.32\textwidth]{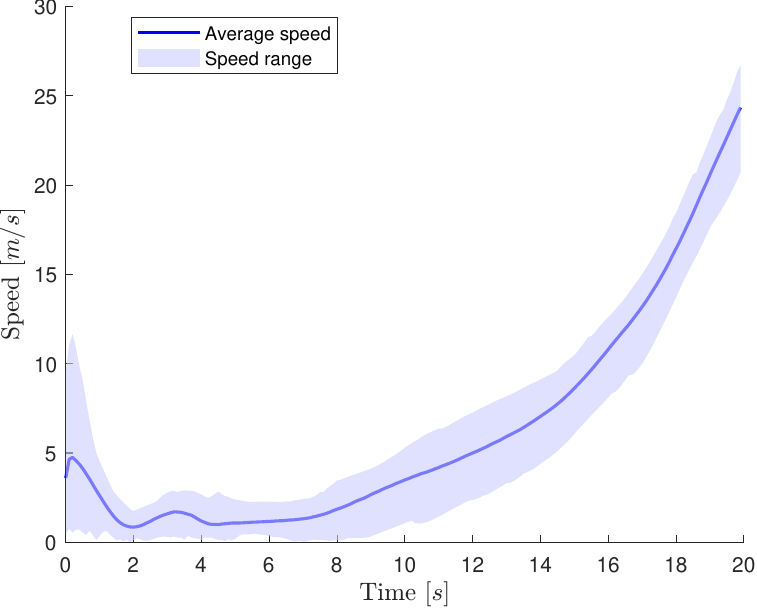}\\
      \includegraphics[width=0.32\textwidth]{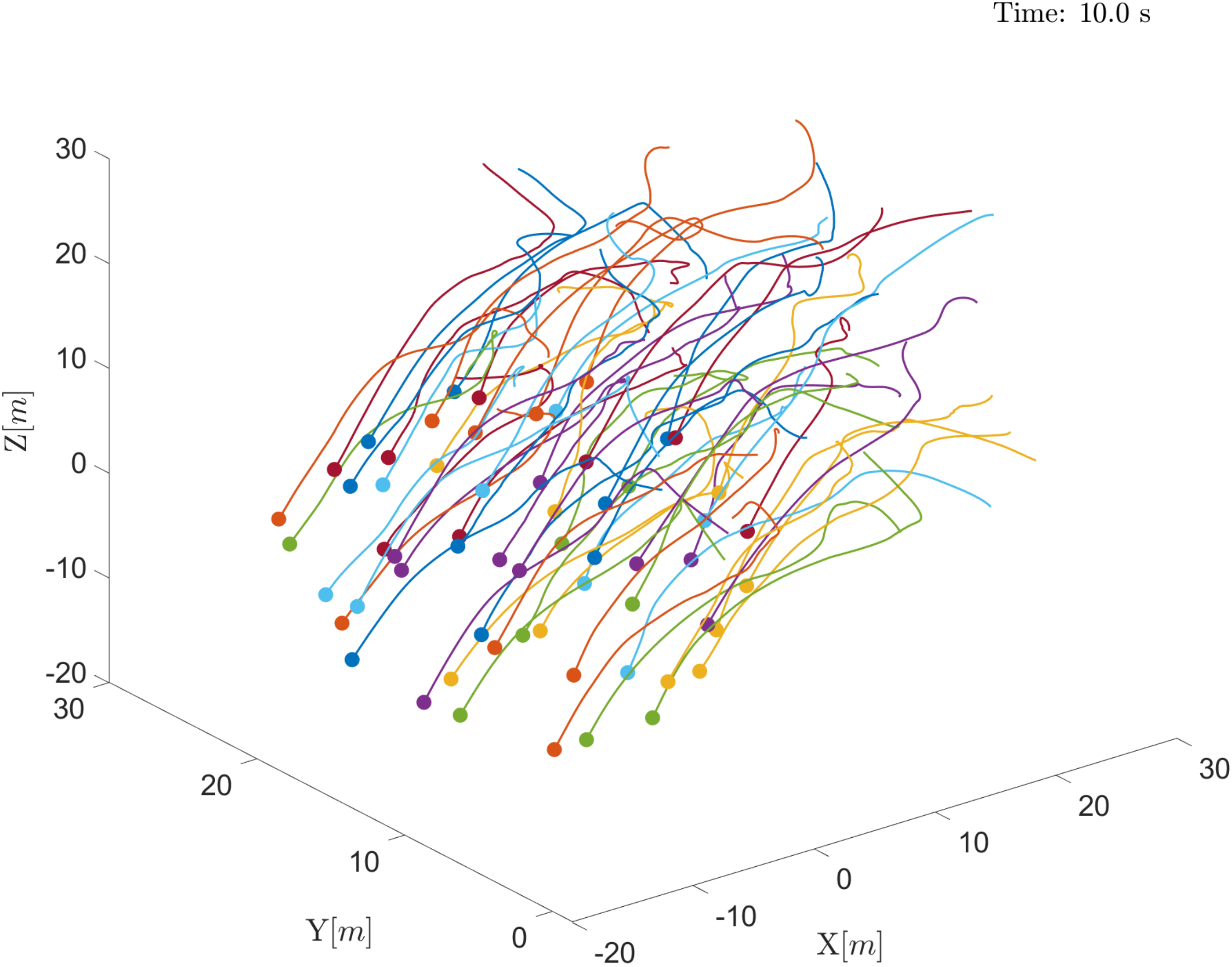}
      \includegraphics[width=0.32\textwidth]{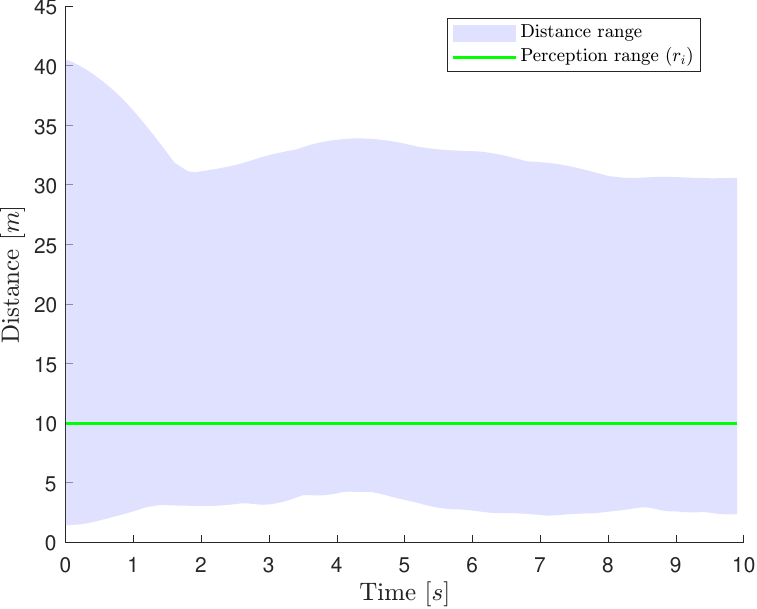}
      \includegraphics[width=0.32\textwidth]{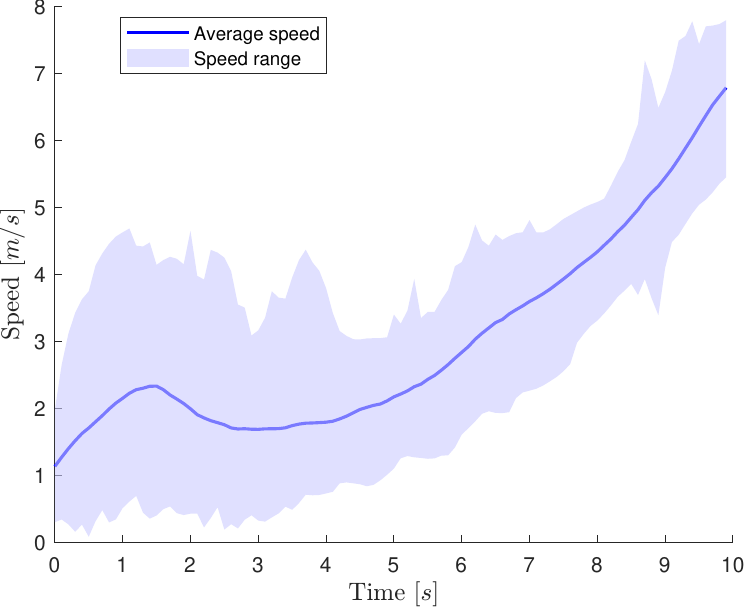}
   \caption{Flocking behavior of \( 60 \) agents for both 2D (top row) and 3D (bottom row) scenarios using the minimal flocking controller. The cohesive and collision-free flocking behavior is successfully established. Inter-agent distances stabilize while maintaining sufficiently safe separation. The speed profiles of the flock agents indicate achieving alignment.}  
   \label{fig:simplified}
\end{figure*}

We also evaluate the minimal flocking model introduced in Section~\ref{sec:simple}. This assessment includes both 2D and 3D flocking scenarios, each involving \(n=60\) agents employing the flocking controller in~\eqref{eq:simple-con}. For each agent \(i\), the radius of perception is defined as \(r_i = 10 \) \(\mathrm{m}\). The agents' positions and velocities were initialized randomly within the range $(0,30)$ \(\mathrm{m}\) and \([-1,1]\) \(\mathrm{m/s}\). The scaling parameters for both 2D and 3D scenarios are set to \(\beta_i = \frac{1}{r_i}\) and \(\sigma_{ij}=\mu_{ij}=1\) for all \((j,i)\in\mathcal{E}\). The total simulation time is $20.0$ \(\mathrm{s}\) for the 2D case and \(10.0\) \(\mathrm{s}\) for the 3D case, with a time step of $0.1$ \(\mathrm{s}\). The simulation results, presented in Fig.~\ref{fig:simplified}, indicate that the agents successfully achieve cohesive and collision-free flocking behavior. This is evidenced by stabilized inter-agent distances that ensure sufficient separation to avoid collisions. Additionally, the speed profiles demonstrate that the agents align effectively. These results emphasize the minimal flocking model's effectiveness in establishing and maintaining a cohesive and aligned flock.

\section{Conclusions}\label{sec:con}

We introduced a distributed perception-based flocking control model that leverages zone-based interactions to achieve adaptable and dynamic flocking behavior without relying on communication or global positioning. By incorporating a conflict zone where repulsive and attractive forces interact, our model captures the fluidity and adaptability of natural flocking. Simulation results validated the model's ability to create and manage a cohesive flock in complex scenarios, such as navigating confined spaces and interacting with alien agents. The model's reliance on local sensing without communication highlights its potential for scalable applications in real-world autonomous systems. Toward less predictable and more lifelike motions, like in realistic flocking, stochastic elements, such as stochastic weights or random perturbations in velocities, can be integrated into the model.

\section*{Acknowledgments} This work was funded by the Czech Science Foundation (GAČR) under research project no. \(\mathrm{23-07517S}\) and the European Union under the project Robotics and Advanced Industrial Production (reg. no. \(\mathrm{CZ.02.01.01/00/22\_008/0004590}\)).

\bibliographystyle{IEEEtran}
\bibliography{mybibliography}

%\begin{thebibliography}{10}
%\end{thebibliography}

\end{document}